\documentclass[aps,prb,reprint,superscriptaddress,amsfonts,amssymb,amsmath]{revtex4-2}

\usepackage{amssymb}
\usepackage{fullpage}
\usepackage{graphicx}
\usepackage{bibunits}
\usepackage{wrapfig}
\usepackage{floatflt}
\usepackage{color}
\usepackage{xspace}
\usepackage{dsfont}
\usepackage[official,right]{eurosym}
\usepackage[top=0.8in, bottom=0.8in, left=0.75in, right=0.75in]{geometry}
\usepackage[utf8]{inputenc}
\usepackage[colorlinks=true,allcolors=black]{hyperref}

\newcommand{\PHCX} {(C$_4$H$_{12}$N$_2$)Cu$_2$(Cl$_{1-x}$Br$_{x}$)$_6$ \xspace}

\newcommand{\SULX} {H$_8$C$_4$SO$_2\cdot$Cu$_2$(Cl$_{1-x}$Br$_x$)$_4$ \xspace}

\newcommand{\DTNX} {Ni(Cl$_{1-x}$Br$_x$)$_2$$\cdot$4SC(NH$_2$)$_2$\xspace}

\newcommand{\dd}{\mathrm{d}}

\newcommand{\be}{\begin{equation} }
\newcommand{\ee}{\end{equation} }
\newcommand{\bea}{\begin{eqnarray} }
\newcommand{\eea}{\end{eqnarray} }

\begin{document}

\title{Finite-temperature dynamics and the role of disorder in nearly-critical \DTNX}

\author{L. Facheris}
\email{lfacheri@phys.ethz.ch}
\affiliation{Laboratory for Solid State Physics, ETH Zürich, 8093 Zürich, Switzerland}
\author{D. Blosser}
\affiliation{Laboratory for Solid State Physics, ETH Zürich, 8093 Zürich, Switzerland}
\author{K. Yu. Povarov}
\affiliation{Laboratory for Solid State Physics, ETH Zürich, 8093 Zürich, Switzerland}
\author{R. Bewley}
\affiliation{ISIS Facility, Rutherford Appleton Laboratory, Chilton, Didcot, Oxon OX11 0QX, United Kingdom}
\author{S. Gvasaliya}
\affiliation{Laboratory for Solid State Physics, ETH Zürich, 8093 Zürich, Switzerland}
\author{A. Zheludev}
\email{zhelud@ethz.ch}
\homepage{http://www.neutron.ethz.ch/}
\affiliation{Laboratory for Solid State Physics, ETH Zürich, 8093 Zürich, Switzerland}

\date{\today}

\begin{abstract}

Inelastic neutron scattering is used to investigate the temperature dependence of spin correlations in the 3-dimensional XY antiferromagnet \DTNX, $ x = 0.14(1)$, tuned close to the chemical-composition-induced soft-mode transition. The local dynamic structure factor shows $\hbar\omega/T$ scaling behavior characteristic of a quantum critical point. The deviation of the measured critical exponent from spin wave theoretical expectations are attributed to disorder. Another effect of disorder is local excitations above the magnon band. Their energy, structure factor and temperature dependence are well explained by simple strong-bond dimers associated with Br-impurity sites.
\end{abstract}

\maketitle

\section{Introduction}

Magnetic insulators make excellent model systems for studying quantum phase transitions. Not only do they often host good realizations of numerous different quantum critical points (QCPs), but they are also particularly amenable to quantitative experimental methods such as neutron spectroscopy. The most extensively studied are magnetic field driven QCPs \cite{GiamarchiRueggTchernyshyov2008,ZapfJaimeBatista2014,Zheludev2020}. In rare cases a different class of QCP is induced by hydrostatic pressure, where small distortions of the crystal structure lead to a continuous modification of magnetic superexchange interactions \cite{TanakaGotoFujisawa2003,RueggFurrerSheptyakov2004,ThedeMannigMansson2014,PerrenMollerHuvonen2015,KuritaTanaka2016,HayashidaZaharkoKurita2018}.
The need for bulky pressure cells makes such transitions difficult to study experimentally. One way around are similar phase transitions induced by chemical substitution in solid solutions, which can also be thought of as ``chemical pressure'' \cite{PovarovMannigPerren2017,HayashidaStoppelYan2019}. The latter case is particularly interesting but also complicated, since the necessary presence of chemical disorder may introduce randomness into the spin Hamiltonian. This, in turn, may have consequences for critical behavior.

The subject of the present study is the well-known 3-dimensional $S=1$ system \DTNX (DTNX for short) \cite{YuYinSullivan2012}. The parent $x=0$ material has a singlet ground state and a spin gap due to strong easy-plane magnetic anisotropy \cite{PaduanFilhoChiricoJoung1981,ZapfZoccoHansen2006,ZvyaginWosnitzaBatista2007,YinXiaZapf2008,MukhopadhyayKlanjsekGrbic2012,TsyrulinBatistaZapf2013,WulfHuvonenSchonemann2015}. Upon Br substitution on non-magnetic Cl sites, the spin gap is suppressed \cite{YuYinSullivan2012,PovarovWulfHuvonen2015}. Beyond a certain critical Br concentration the system exhibits long-range antiferromagnetic order, while the excitation spectrum is gapless with a linear dispersion \cite{PovarovMannigPerren2017,MannigPovarovOllivier2018}. Two key questions regarding this chemical composition induced transition remain: 1) Is this indeed a true QCP with quantum-critical fluctuation dynamics subject to scaling laws? and 2) How important are the inherent disorder and randomness in the magnetic Hamiltonian for this transition? The two issues are closely connected. In \DTNX the Harris criterion \cite{Harris1974} is violated, meaning disorder must be relevant. Moreover, in a 2-dimensional Heisenberg model that is in many ways similar to that for DTNX the QCP is not even expected to survive disorder \cite{Vojta2013}.

Below we report measurements on an almost critical \DTNX sample using thermodynamic and neutron scattering experiments in a wide range of temperatures. We find that at low temperatures the excitations remain well described by the same RPA model as used to understand the parent compound. The linear portion of the spectrum shows a finite-$T$ scaling behavior over at least one and a half decades in $\hbar\omega/T$. Disorder is not entirely irrelevant though. It may be responsible for the observed deviation of the scaling exponents from those expected for the disorder-free QCP. In addition, it generates local excitations at high energies \cite{PovarovWulfHuvonen2015}. We clarify the microscopic origin of the latter through a measurement of their structure factor and temperature dependence.

\section{Material and Methods}

DTNX shares many similarities with its parent compound DTN. A detailed description of the structure and interactions can be found, for example, in Ref.~\cite{MannigPovarovOllivier2018}. The material is tetragonal body-centered, space group $I4$ \cite{YuYinSullivan2012}. Magnetism originates from $S=1$ Ni$^{2+}$ ions that form chains along the crystallographic $c$ axis. Strong easy-$(a,b)$-plane single-ion anisotropy produces a spin-singlet ground state with $S^z=0$ and an energy gap $\Delta\sim 0.3$~meV to the lowest-energy $S^z=\pm 1$ excitation doublet. Other relevant interactions are the intra-chain antiferromagnetic (AF) Heisenberg exchange interaction $J_c\sim 0.17$~meV and the weak inter-chain coupling $J_a\sim 0.013$~meV \cite{TsyrulinBatistaZapf2013}.

Br substitution does not appreciably alter the lattice parameters but affects the spin Hamiltonian and magnetic properties \cite{YuYinSullivan2012}. As the concentration $x$ in \DTNX is increased the gap decreases from $\Delta = 0.3$~meV in DTN \cite{PaduanFilhoGratensOliveira2004}, to 0.2~meV for $x = 0.06$ \cite{PovarovWulfHuvonen2015}. The critical concentration at which the spin gap closes was estimated to be around $x_c = 0.16$ \cite{PovarovMannigPerren2017}. For $x = 0.21$ the system is well into the long range ordered phase with a gapless linear spin wave spectrum \cite{PovarovMannigPerren2017}.

For the present study, fully deuterated DTNX single crystals were grown from aqueous solution. We aimed at approaching the critical concentration as close as possible. After numerous attempts the best sample was chosen with a bromine content $x = 0.14(1)$. The latter was confirmed in single crystal X-ray diffraction measurements on an APEX-II Bruker instrument. 

Thermodynamic data were collected using 1.0(5)~mg-size single crystals on a commercial Quantum Design Physical Properties Measurement System (PPMS) with a dilution cryostat insert and a $14$~T superconducting magnet. The magnetic field was in all cases applied along the crystallographic $c$ axis.

Neutron scattering was performed at the high-resolution cold-neutron TOF spectrometer LET at the ISIS Neutron facility. We employed two co-aligned samples of total mass $\sim 1.2$~g. Sample mosaic was about $1.5^\circ$ full width at half height (FWHM). The sample was mounted with the $(h,h,l)$ reciprocal-space plane horizontal. Sample environment was a $^3$He cryostat with a base temperature $T = 300$~mK. The data were collected in several frames simultaneously with incoming neutron energies $E_i = 1.45$, $2.20$ and $3.71$~meV, respectively. The measured energy resolution at the elastic position was $31~\mu$eV, $55~\mu$eV and $111~\mu$eV FWHM for the three configurations, correspondingly. The data were collected while rotating the sample in $1^\circ$ steps around the vertical $(1,-1,0)$ axis, covering a total angle of slightly over $50^\circ$. Typical counting time at each step was 15~min.

\section{Results and discussion}

\subsection{Calorimetry}

\begin{figure}
\includegraphics[width=\columnwidth]{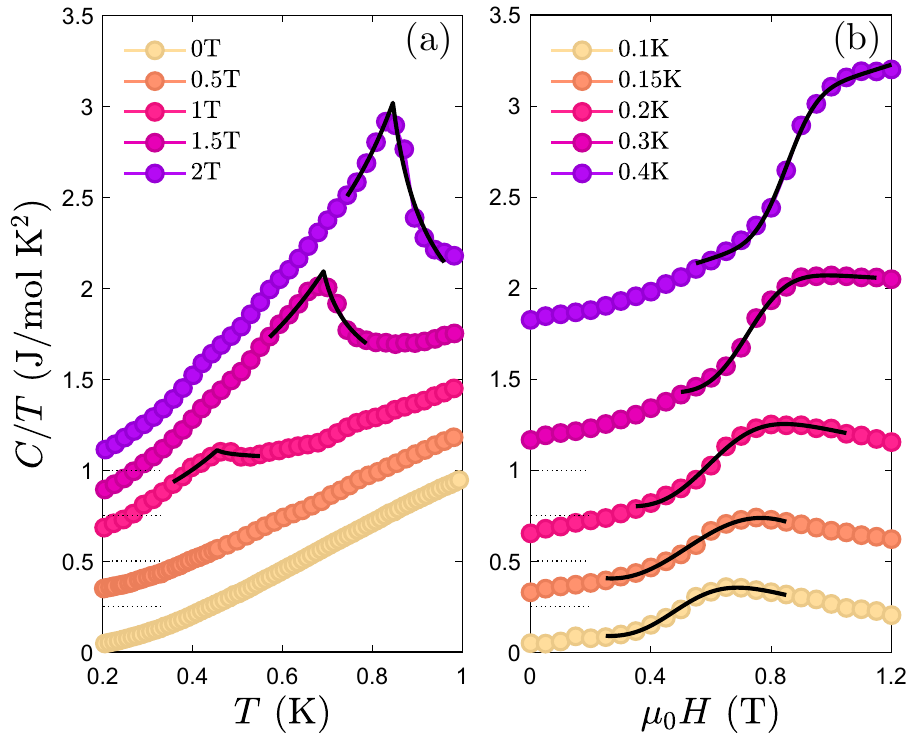}
\caption{\label{Fig-specheat} Specific heat measured in DTNX, $x=0.14(1)$, in magnetic fields applied along the crystallographic $c$ axis using constant-$H$ (a) and constant-$T$ (b) scans. Solid lines are empirical fits to the data to pinpoint the transition, as described in the text. For visibility, the scans are offset by $0.25$~J/mol K$^2$ relative to one another.}
\end{figure}
	
\begin{figure}
	\includegraphics[width=\columnwidth]{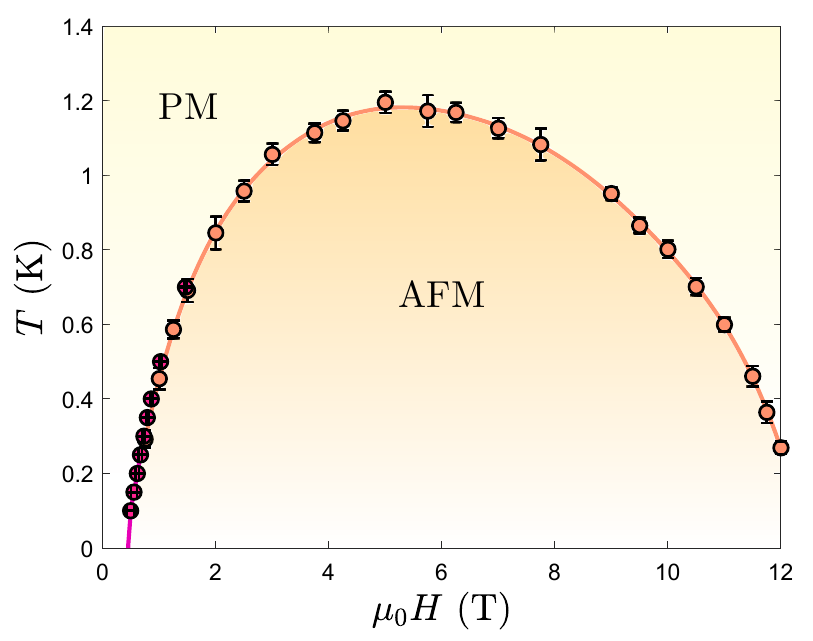}
	\caption{\label{Fig-phase} Magnetic phase diagram of DTNX, $x=0.14(1)$, in magnetic fields applied along the crystallographic $c$, as derived from thermodynamic measurements. The solid line is a guide for the eye.}
\end{figure}

The proximity of our DTNX sample to the critical point was quantified by thermodynamic measurements.
Typical heat capacity data collected as a function of temperature and field are shown in Figs.~\ref{Fig-specheat}(a) and \ref{Fig-specheat}(b), respectively. No magnetic ordering transition is observed in zero applied field, suggesting that the sample is slightly ``underdoped''. In fields exceeding $\mu_0H = 1$~T there is a clear lambda anomaly in the measured temperature dependencies, which we attribute to the onset of long range magnetic order. The transition temperatures were determined by empirical power-law fits in the vicinity of the peak [solid lines in Fig.~\ref{Fig-specheat}(a)], similarly to how it was done in Ref.~\cite{HuvonenZhaoMansson2012}. At the lowest temperatures the transition is associated with an inflection point in the $C(H)$ curve. The corresponding transition field was pinpointed by empirical error function fits [solid lines in Fig.~\ref{Fig-specheat}(b)]. The phase boundary combined from both types of measurements is shown in Fig \ref{Fig-phase}. A linear extrapolation \cite{WulfHuvonenSchonemann2015} to $T=0$ provides an estimate of the residual gap energy: $\Delta =g\mu_BH_c = 0.059(5)$~meV assuming $g=2.26$ \cite{ZvyaginWosnitzaBatista2007}. The sample is thus indeed close to the critical Br concentration at which the gap closes in the absence of any external magnetic field. We note however that the observed specific heat anomalies, particularly those in field scans at a constant temperature, are considerably less pronounced than in the parent DTN compound \cite{KohamaSologubenkoDilley2011,WulfHuvonenSchonemann2015}. This effect is undoubtedly due to the halogen-site disorder in the sample, as had been observed in several other systems such as \SULX \cite{WulfMuhlbauerYankova2011} and \PHCX \cite{HuvonenZhaoMansson2012}.

\subsection{Neutron spectroscopy}
\subsubsection{Magnon dispersion at base temperature}

\begin{figure*}
\includegraphics[width=\textwidth]{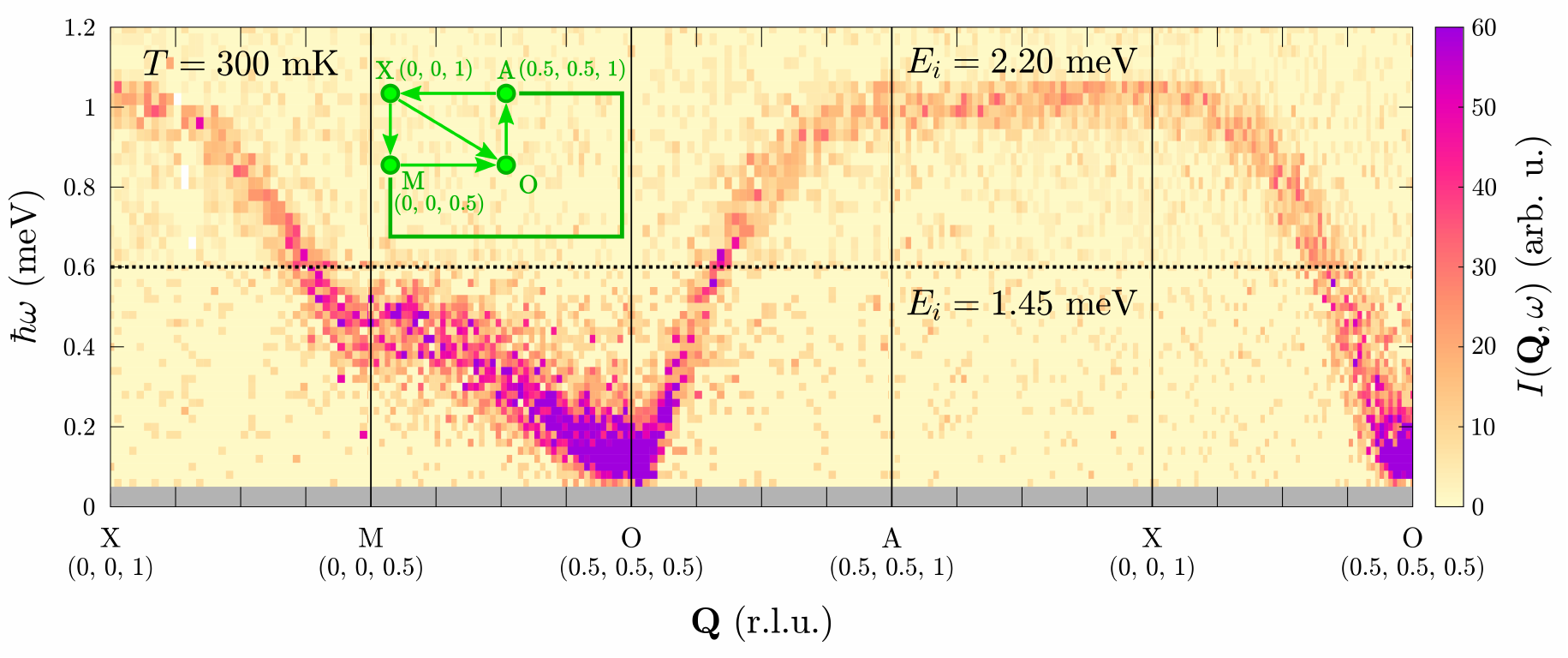}
\caption{\label{Fig-spectrum} False color plot of the neutron scattering intensity measured \DTNX, $x=0.14(1)$, at $300$~mK. Several slices along high-symmetry reciprocal space directions are shown. The depth of each slice is $\pm 0.02$~r.l.u. in the transverse reciprocal space directions. The corresponding trajectories are shown in the inset. The data above the dotted divider line were collected using neutrons with $E_i = 2.20$~meV. The low-energy data were measured with $E_i = 1.45$~meV. }
\end{figure*}

A false-color plot of the inelastic intensity measured at the lowest temperature $T = 300$~mK is shown in Fig.~\ref{Fig-spectrum}. It represents a combination of cuts along different reciprocal-space trajectories, as shown in the inset. The depth of each cut is $\pm0.02$~r.l.u. in both perpendicular reciprocal-space directions. A constant background was estimated away from the obvious magnon branch and subtracted. The magnetic scattering intensity is directly proportional to the dynamic structure factor $\mathcal{S}(\mathbf{q},\omega)$, i.e., to the spatial and temporal Fourier transform of the spin-spin correlation function \cite{Squiresbook1978}. $\mathcal{S}(\mathbf{q},\omega)$ was obtained by correcting the background-subtracted intensity for the magnetic form factor of Ni$^{2+}$ ions and for the neutron polarization factor, assuming scattering to be purely magnetic and polarized transverse to the magnetic easy-axis \cite{MannigPovarovOllivier2018}. The data measured with two different incident neutron energies are combined with relative scaling based on the intensity of the quasielastic line. The spectrum appears qualitatively similar to the one measured in Refs.~\cite{PovarovWulfHuvonen2015,MannigPovarovOllivier2018} for underdoped and overdoped samples, respectively. 

A look at Fig.~\ref{Fig-spectrum} suggests that the gap may actually be considerably larger than estimated from thermodynamics. This, however, is an illusion due to finite wave vector resolution. The latter is in part instrumental and in part due to the chosen binning in reciprocal space. Due to finite resolution, near the sharp dispersion minima at the AF zone-centers, the intensity maximum will always be {\em above} the actual gap energy. For DTN and strongly ``underdoped'' DTNX \cite{PovarovWulfHuvonen2015} the gap is larger, the dispersion minimum less sharp, and the resolution effects pose much less of a problem. In our case of an almost-critical sample they are quite severe. 

\begin{figure}
	\includegraphics[width=\columnwidth]{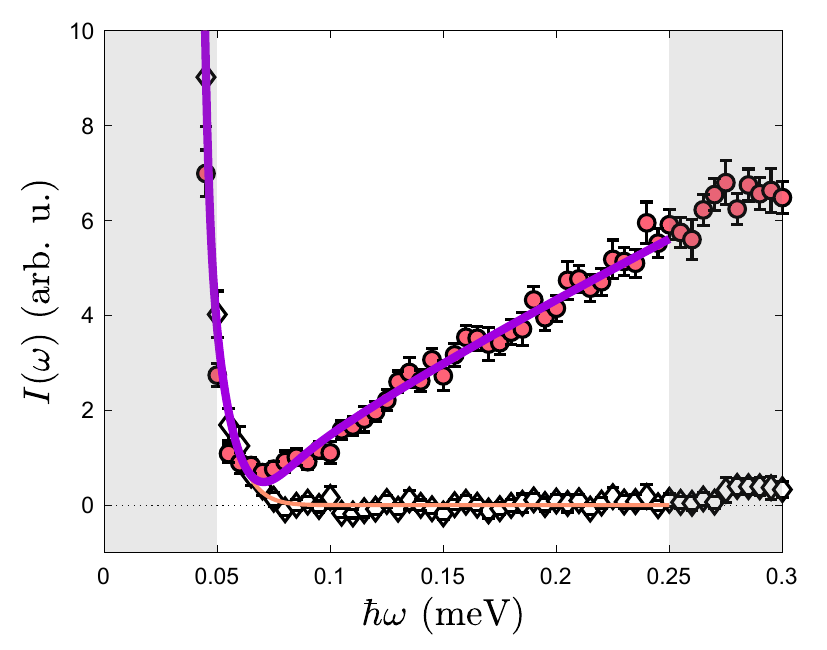}
	\caption{\label{Fig-int} Wave-vector integrated intensity measured near the magnon dispersion minimum $(-0.5,-0.5,-1.5)$ in \DTNX, $x=0.14(1)$, at $T=300$~mK (solid symbols).The wave vector integration range is $-0.75<h<-0.25$, $-1.75<l<-1.25$ in the $(h,h,l)$ plane and $-0.25<h<0.25$ along the vertical $(h,-h,0)$-direction. The background (open symbols) was measured away from the AF zone-center. The lines are fits as described in the text. The data in shaded areas was excluded from the fit.}
\end{figure}

A way around the problem is to integrate over momentum transfer and thus compute the local dynamic structure factor $\mathcal{S}(\omega)$. We performed such integration for the vicinity of the AF zone-center at $(-0.5,-0.5,-1.5)$. The integration range was $-0.75<h<-0.25$, $-1.75<l<-1.25$ in the $(h,h,l)$ plane and $-0.25<h<0.25$ along the vertical $(h,-h,0)$-direction. The result is shown in Fig.~\ref{Fig-int} in solid symbols.
The background (Fig.~\ref{Fig-int}, open symbols) was obtained near the maximum of the $c$-axis dispersion, with the same integration range in the $(h,h,0)$ and $(h,-h,0)$ directions and for $-2<l<-1.75$ and $-1.25<l<-1$. 

At low temperatures, magnetic excitations in DTNX have been very successfully described using Generalized Spin Wave Theory/ Random Phase approximation (GSWT/RPA) \cite{MatsumotoKoga2007,ZhangWierschemYap2013,MunizKatoYasuyuki2014} for all Br concentrations \cite{MannigPovarovOllivier2018}. In that theory the magnon intensity is inversely proportional to energy. Near the AF zone-center the dispersion has a ``relativistic'' form
\be
(\hbar \omega_\mathbf{q})^2=(q_a^2+q_b^2)c_\bot^2+q_c^2c_\|^2+\Delta^2,
\ee
where the three components of the wave vector $\mathbf{q}$ are measured relative to the AF zone-center, and $c_\bot$ and $c_\|$ are spin wave velocities in the $(a,b)$ plane and along the $c$-axis, respectively. The local dynamic structure factor is then obtained by integrating out the momentum transfer:
\be
\mathcal{S}(\omega)\propto\int \dd^3\mathbf{q} \,\frac{\delta(\hbar\omega-\hbar\omega_\mathbf{q})}{\hbar\omega_\mathbf{q}}\propto\frac{(\hbar\omega)^2-\Delta^2}{\hbar\omega}.\label{intint}
\ee
To estimate the spin gap in our sample, we first fit the background to an empirical sum of two Gaussians (thin line in Fig.~\ref{Fig-int}). We combined the fitted background with Eq.~\eqref{intint}, the latter convoluted with the measured energy resolution of the instrument. The resulting function was used to analyze the measured integrated intensity (solid symbols) in the range $0.05$~meV $<\hbar \omega< 0.25$~meV. A least squares fit of an overall scale factor for Eq.~\eqref{intint} and the gap energy yields $\Delta=0.06(1)$~meV, in excellent agreement with the thermodynamic estimate. The fitted curve is shown in a solid line in Fig.~\ref{Fig-int}.

\begin{figure}
	\includegraphics[width=\columnwidth]{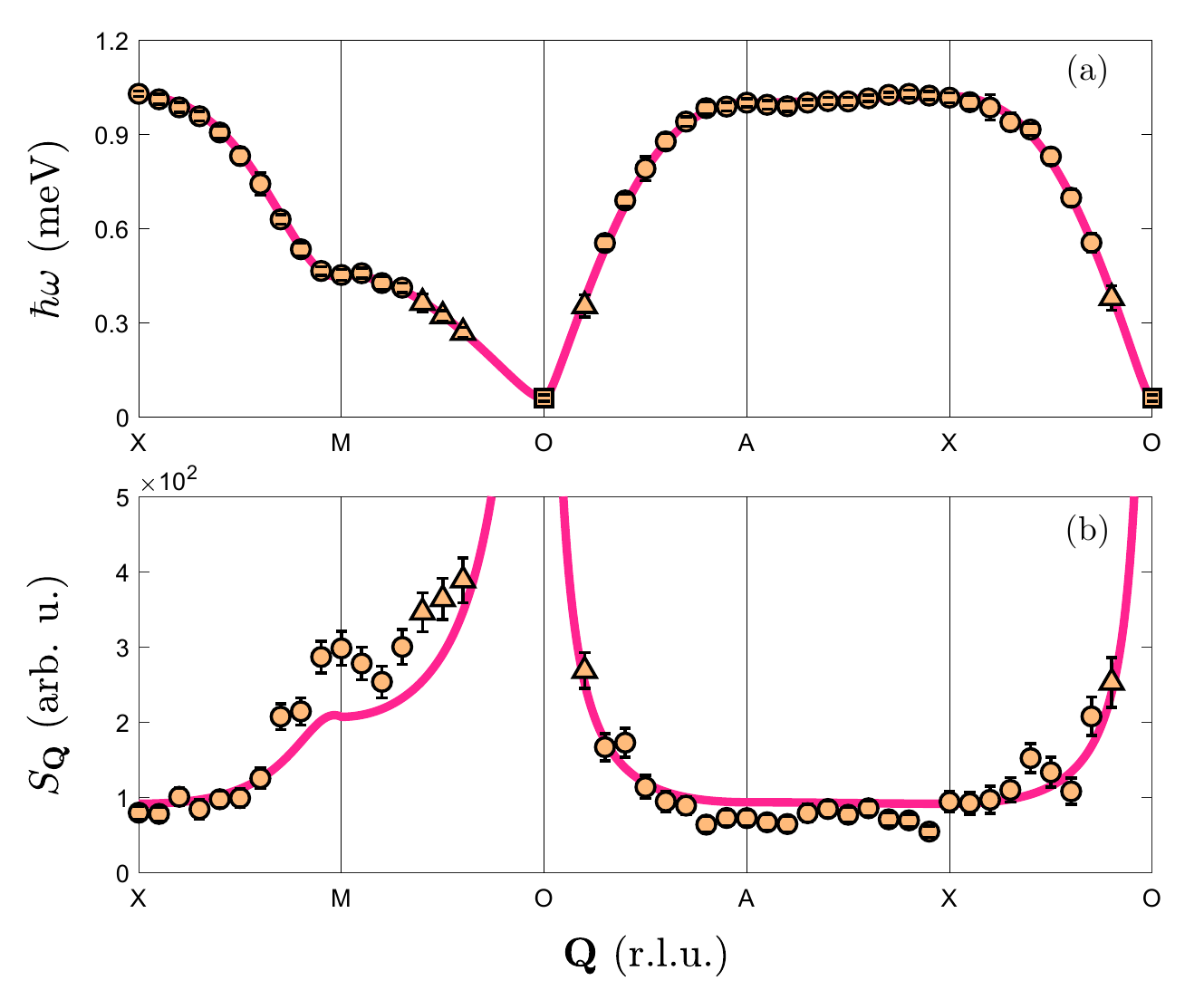}
	\caption{\label{Fig-disp} (a) Magnon dispersion in DTNX, $x=0.14(1)$, as obtained from Voigt fits to constant-$\bm{Q}$ cuts through the inelastic neutron data (see text). (b) The corresponding energy-integrated intensities. Circles and triangles refer to measurements with $E_i=2.20$~meV and $E_i=1.45$~meV, respectively. The square represents the magnon gap as determined from the analysis of $\mathcal{S}(\omega)$. Solid lines are fits to a GSWT/RPA-based model, as described in the text.}
\end{figure}

Away from the zone-center, resolution effects are less important and we can perform an analysis similar to that done in Ref. \cite{PovarovWulfHuvonen2015}. The measured intensity data were binned into constant-$\mathbf{Q}$ cuts lined up along the high-symmetry reciprocal space trajectories shown in Fig.~\ref{Fig-spectrum}. The typical bin size was $0.05 \times 0.05 \times 0.05$~r.l.u. along the $(h,h,0)$, $(h,-h,0)$ and $(0,0,l)$ directions, respectively. These ``scans'' were fit to Voigt functions. The peak position, Lorentzian width, an intensity prefactor and a constant background were the fitting parameters. The Gaussian component was fixed to the calculated energy resolution of the instrument combined with an additional broadening due to wave vector binning (estimated from the bin size and dispersion slope). The fitted peak positions are plotted versus wave vector in Fig.~\ref{Fig-disp}(a). Fig.~\ref{Fig-disp}(b) shows the corresponding intensities. The fitted intrinsic excitation widths were in all cases smaller than 0.12~meV.

The dispersion relation measured above 0.25~meV energy transfer was analyzed using the same model Hamiltonian as previously adopted for the $x=0.06$ material \cite{PovarovWulfHuvonen2015}. The parameters of the model are the single-ion anisotropy $D$; the nearest and next-nearest neighbor exchange constants along the $c$ axis $J_c$ and $J_{c2}$; coupling along the crystallographic $a$ axis $J_a$ and a diagonal coupling connecting to nearest-neighbor Ni$^{2+}$ sites from the two inter-collated Bravais lattices in the body-centered structure of DTNX \cite{PovarovWulfHuvonen2015,TsyrulinBatistaZapf2013}. The dispersion was calculated within the GSWT/RPA framework as in Ref.~\cite{PovarovWulfHuvonen2015}. The parameters were at all times constrained to give the gap energy as estimated from the local structure factor. Least-squares Levenberg-Marquardt fitting with initial parameters from Ref.~\cite{PovarovWulfHuvonen2015} showed rapid and robust convergence. The result of the fit is plotted in a solid line in Fig.~\ref{Fig-disp}(a). The fitted parameter values and variances are tabulated in Table~\ref{Parameter_values}. Note that the relative variances are very small, on the level of systematic inaccuracies such as the spectrometer calibration. From a direct comparison with parameters previously determined for $x=0.06$ it becomes obvious that the two-fold reduction of the spin gap results from a very subtle change in the balance between single ion anisotropy and $c$-axis exchange interactions, particularly the $D/J_c$ ratio.

\begin{table}
\begin{ruledtabular}
\begin{tabular}{ccc}
 & $x = 0.06$ (Ref.~\cite{PovarovWulfHuvonen2015}) & $x = 0.14$ (present work)\\
\hline
$D$ & 0.807(2) & 0.81(2) \\
$J_c$ & 0.150(1) & 0.153(3) \\
$J_a$ & 0.0157(2) & 0.0155(2) \\
$J_d$ & 0.0060(4) & 0.0061(3) \\
$J_{c2}$ & -0.0096(7) & -0.018(1)\\
\hline
$\Delta$ & 0.2 & 0.06(1)\\
$D/J_c$ &   5.38(3) &  5.29(2)
\end{tabular}
\end{ruledtabular}
\caption{\label{Parameter_values} Fitted spin Hamiltonian parameters for DTNX, $x=0.14 (1)$ in comparison with those for $x=0.06$ \cite{PovarovWulfHuvonen2015}. All energies are given in meV units.}
\end{table}

\subsubsection{Temperature dependence}
An important result of this work pertains to the temperature dependence of magnetic excitations in DTNX. At a QCP and inside the quantum-critical regime we expect the critical spin fluctuations at low energies to obey scaling laws \cite{SachdevBook2011,SachdevKeimer2011,Zheludev2020}. Temperature itself becomes the only relevant energy scale. In particular, the local (momentum-integrated) dynamic structure factor $\mathcal{S}(\omega)$ can be expressed through a scaling function of $\hbar \omega/k_BT$:
\bea
\mathcal{S}(\omega) \propto (k_BT)^{a}\Phi\bigg(\frac{\hbar\omega}{k_BT}\bigg), \mathrm{or}\\
\mathcal{S}(\omega) \propto (\hbar \omega)^{a}\mathcal{F}\bigg(\frac{\hbar\omega}{k_BT}\bigg),
\label{ScalingForm}
\eea
with $\mathcal{F}(x)=x^{-a}\Phi(x)$. The exponent $a$ depends on the universality class of the QCP. In the absence of disorder effects, a 3-dimensional XY system with a linear spectrum (dynamical exponent $z=1$) is at the upper critical dimension. Apart from possible logarithmic corrections, we expect mean field behavior, which also describes the GSWT. For the latter, at the point of gap closure the dispersion is indeed linear near the zone center and the magnon intensity is proportional to $1/\omega$, just like for conventional spin waves in the ordered state. Integrating over wave vector yields $a=1$ for that model. 

\begin{figure*}
	\includegraphics[width=\textwidth]{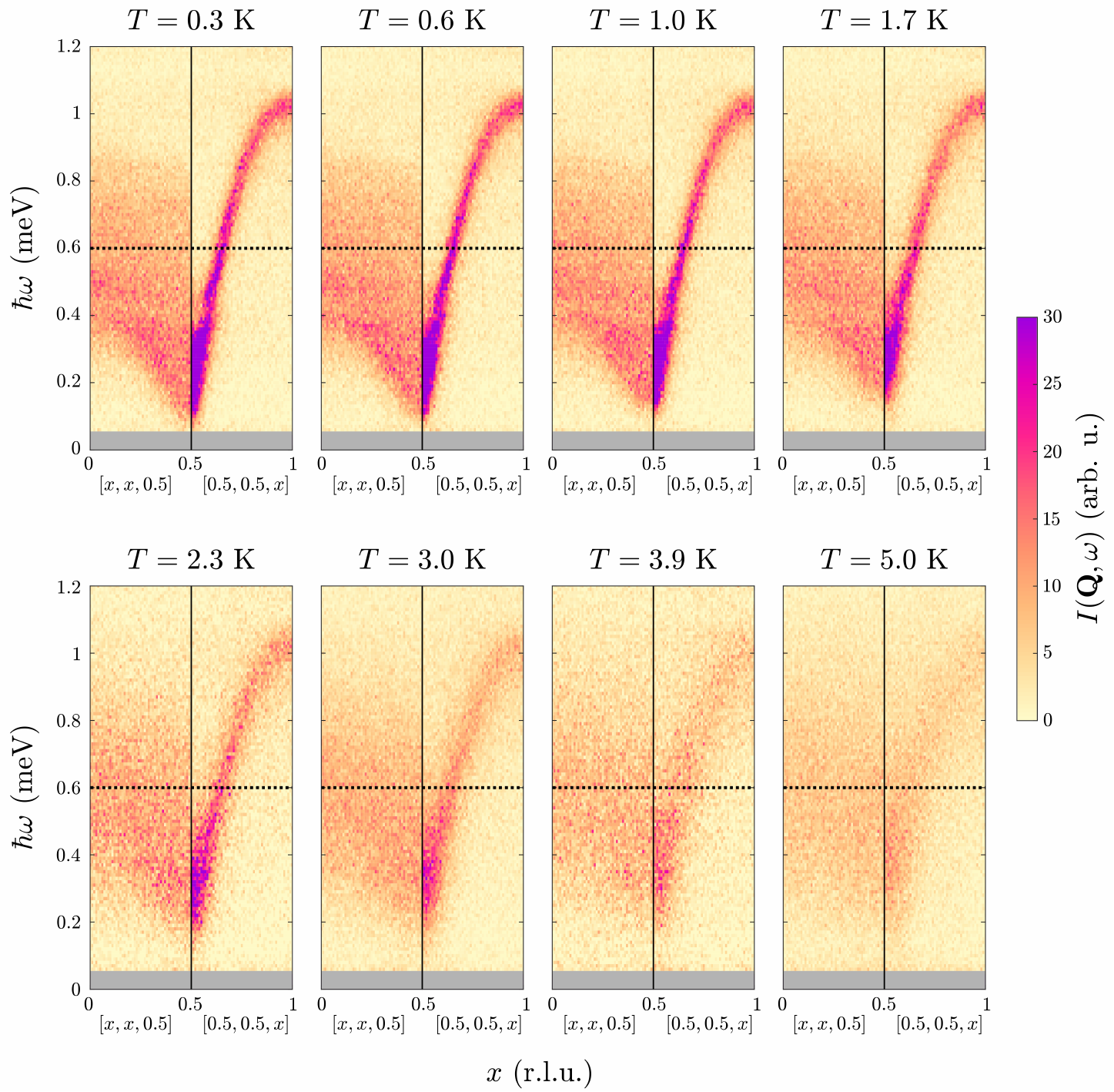}
	\caption{\label{Fig-vart} Neutron scattering intensity measured in \DTNX, $x=0.14(1)$, at several temperatures. The data were integrated in the range $-0.25<h<0.25$ in the $(h,-h,0)$ direction. In the left panel of each graph an integration was also performed in the range $-1.75<l<-1.25$ along the $(0,0,l)$ direction. Instead, the intensity in the right panels was integrated in the range $-0.75<h<-0.25$ along $(h,h,0)$. The dashed line separates data collected with $E_i=2.20$~meV and $E_i=1.45$~meV, respectively.}
\end{figure*}

\begin{figure}
	\includegraphics[width=8.6 cm]{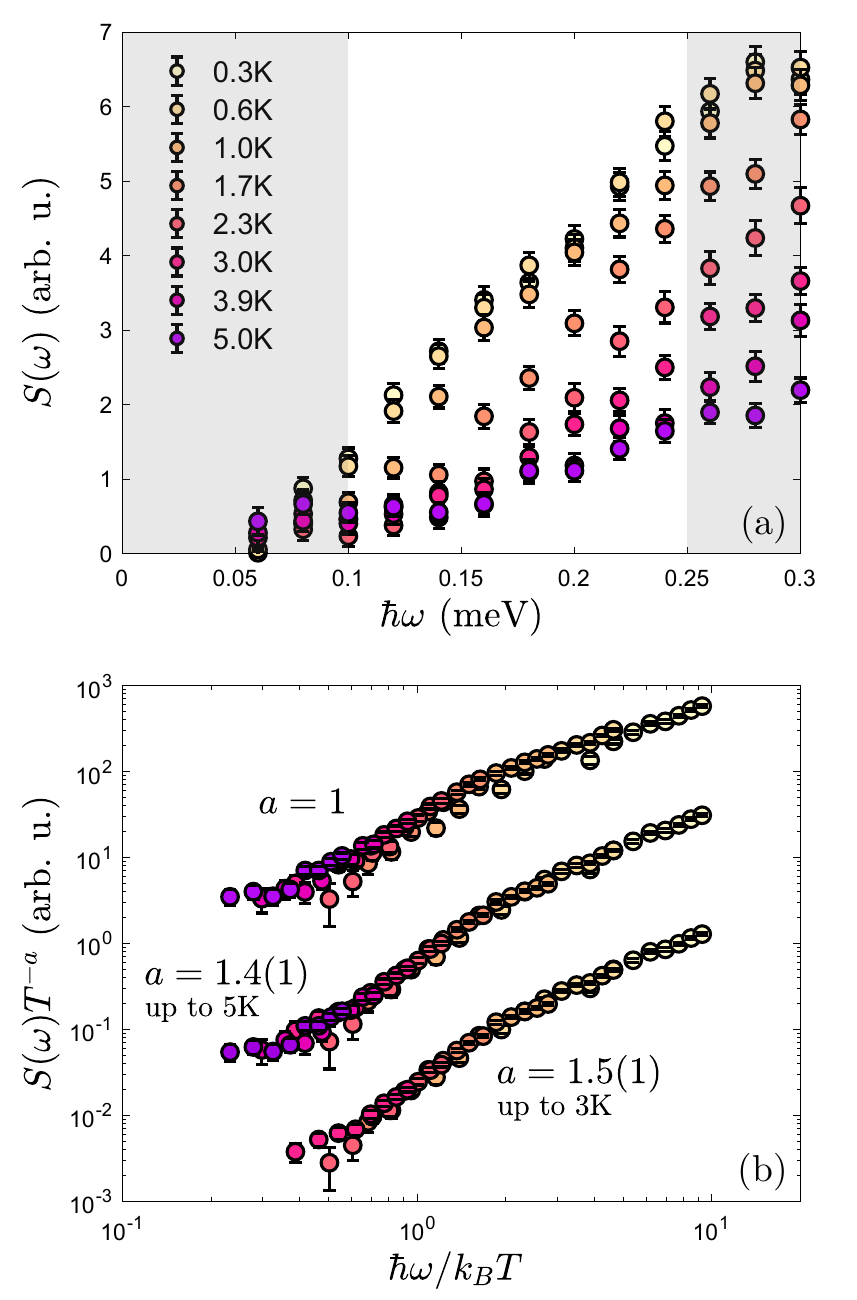}
	\caption{\label{Fig-sq}(a) Local dynamic structure factor measured in \DTNX, $x=0.14(1)$, at different temperatures. The background shown in Fig.~\ref{Fig-int} has been subtracted. Shaded areas mark the boundaries of the region where scaling behavior can be expected to occur. (b) Data between 0.1 and 0.25~meV plotted in scaled variables using a scaling exponent as in GSWT ($a=1$), one that optimizes the collapse of all data up to $T=5$~K ($a=1.4(1)$) and all data up to $T=3$~K ($a=1.5(1)$). Note the logarithmic scale. The plots have been shifted apart vertically for visibility.}
\end{figure}

To see if this prediction applies to DTNX, we studied the temperature dependence of scattering in our sample. Neutron spectra collected at several temperatures are shown in Fig.~\ref{Fig-vart}. Here the intensity has been integrated over a wide range in momentum in two perpendicular directions. The integration range is the same as for the gap estimation above.
Integrating over the third direction and subtracting the background determined at base temperature (see previous section) yields the $\mathcal{S}(\omega)$ plots shown in Fig.~\ref{Fig-sq}(a).

Any scaling behavior can be only expected for data collected in a certain energy range and only at certain temperatures. For a $z = 1$ QCP these should correspond to the width of the approximately linear part of the spectrum. For DTNX this consideration sets an upper bound of $\hbar \omega_\mathrm{max}\approx 0.35$~meV and $T_\mathrm{max}\approx 4$~K, limited by the bandwidth of magnon dispersion along the ``slowest'' M-O reciprocal space direction (Fig.~\ref{Fig-disp}). We chose an upper bound of 0.25 meV instead, to be entirely on the safe side. Regarding temperature, we analyzed the data collected up to 3~K (again on the safe side), and then separately up to 5~K. As a side note, the conditions are usually more stringent in the case of $z=2$ QCPs. There, one has to look for the energy range where the dispersion is well approximated as parabolic. In most cases this constitutes a smaller fraction of the magnon bandwidth \cite{BlosserKestinPovarov2017,BlosserBhartiyaVoneshen2018}. In particular, for the {\em field}-induced $z=2$ transition in DTN, where the magnon bandwidth along the M-O line is under 0.2~meV $\sim 2$~K \cite{ZapfZoccoHansen2006}, numerical studies show that scaling can only be expected to work for $T\lesssim 0.5$~K \cite{Kawashima2004}.

Since our sample is slightly off the critical concentration, one should also avoid the lowest energies \cite{Zheludev2020}. As a conservative choice, we took the lower cutoff at 0.1~meV, roughly twice the gap energy. The energy limits are indicated by the shaded areas in Fig.~\ref{Fig-sq}(a). Similar (and in some cases less stringent) criteria regarding valid energy and temperature ranges have previously been successfully applied to the study of scaling at other $z=1$ quantum critical points \cite{Zheludev2020,HaelgHuvonenGuidi2015,PovarovSchmidigerReynolds2015,HaelgHuvonenButch2015,LakeTennantFrost2005}. 

The scaling plots for our DTNX data in the restricted energy range are shown in Fig.~\ref{Fig-sq}(b). There is no fitting involved in the $a=1$ plot which includes all data up to 5~K. The data collapse for this expected value of the exponent is reasonable, but not quite perfect. Following Ref.~\cite{JeongRonnow2015} we introduced a measure of data collapse based on a 5th-order polynomial fit to the scaled data. The best data collapse is found for $a=1.4(1)$ for all data up to 5~K and $a=1.5(1)$ for data up to 3~K. The resulting scaling plots are also shown in Fig.~\ref{Fig-sq}(b). We have also performed a windowing analysis in energy by fitting only the data within a progressively shrinking energy window, analogously to the procedure outlined in Ref.~\cite{HuvonenZhaoMansson2012}. The conclusion is that shrinking the fit interval produces no statistically significant change in the fitted value of exponent $a$ compared to the value obtained for the full range, which remains within the error bar for each fit.

Of course it can never be fully excluded that the observed deviations from mean field scaling are an artifact of the residual energy gap or the proximity of non-linear part of the spectrum. Nevertheless, with this disclaimer, we can attribute the unusual scaling exponent to the effects of disorder in the system and a resulting re-distribution of the density of states at low energies. We hope that future theoretical work will provide a more concrete interpretation of this result.

\subsubsection{Local excitations}
Previous measurements on the $x=0.06$ compound have revealed the presence of disorder-induced local excitations in DTNX, which appear just above the magnon dispersion maximum \cite{PovarovWulfHuvonen2015}. These local impurity states were also detected by NMR \cite{OrlovaBlinderKermarrec2017}. Theory suggested that they may be viewed as excitations in effectively {\em isolated} $S=1$ dimers composed of nearest neighbor Ni$^{2+}$ ions bound by superexchange interactions substantially enhanced by the Br-impurity between them \cite{DupontCapponiLaflorencie2017}. Being a {\em local} probe, NMR is unable to unambiguously confirm the dimer model. Instead, to verify this simplistic interpretation, we investigated the wave vector and temperature dependence of these excitations in our neutron data.
\begin{figure}
	\includegraphics[width=0.85\columnwidth]{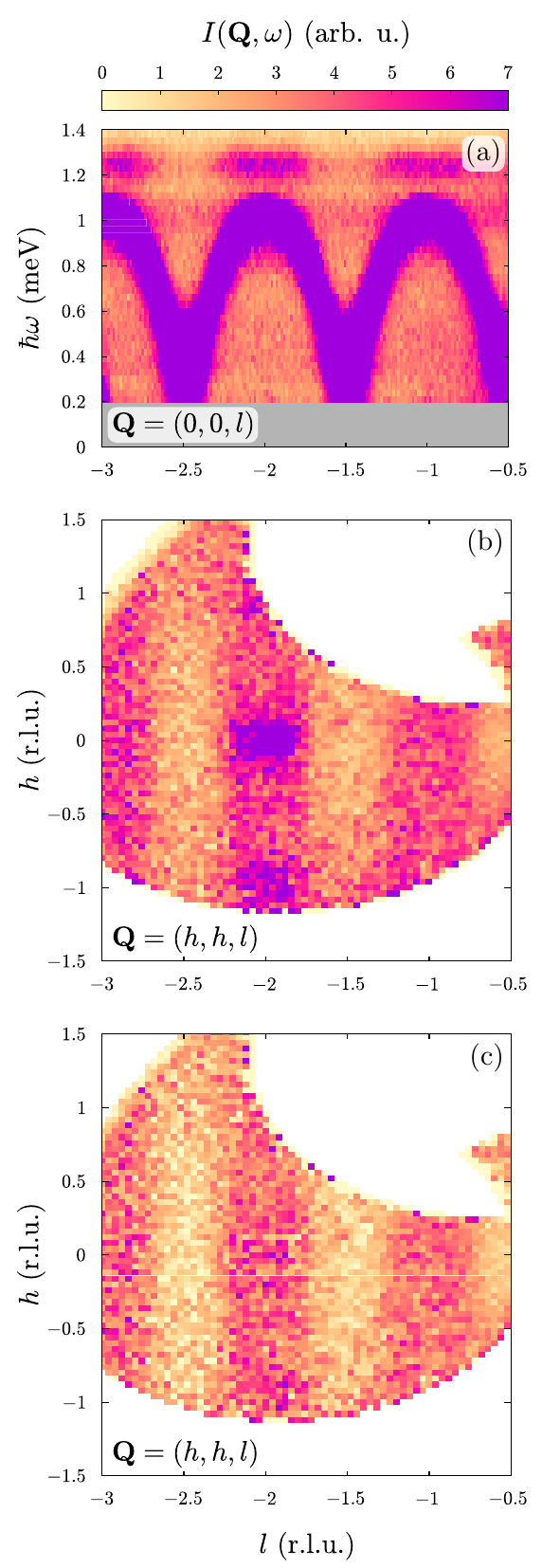}
	\caption{\label{Fig-states} (a) False color plot of the intensity measured in \DTNX, $x=0.14(1)$, in projection onto the crystallographic $c$ axis. The integration is over all the available data in the transverse directions. (b) Constant-energy cut at $\hbar\omega=1.25$~meV integrated
		in the range $\pm 0.1$~meV in energy and $\pm 0.3$~r.l.u. in the $(h,-h,0)$ direction.
		(c) Same as (b), with a background measured at $\hbar\omega=1.45$~meV and subtracted. In all the cases,
		$T=300$~mK and $E_i=3.71$~meV.}
\end{figure}

Similarly to Fig.~7 in Ref.~\cite{PovarovWulfHuvonen2015}, Fig.~\ref{Fig-states}(a) in the present work shows a projection of the measured inelastic scattering intensity integrated in the directions transverse to the $c$ axis over all available data ($E_i=3.71$~meV). The features at $\hbar \omega=1.25$~meV are clearly visible. Fig.~\ref{Fig-states}(b) shows the corresponding constant-energy slice integrated in the range $\pm 0.3$~r.l.u. in the $(h,-h,0)$ direction and $\pm 0.1$~meV in energy. A periodic modulation of intensity of the impurity mode along the $c$ axis is unmistakable in both projections. The apparent additional modulation along the $(h,h,0)$ direction is actually due to scattering by strong phonons emanating from the $(-1,-1,-2)$ and $(0,0,-2)$ Bragg peaks. This phonon background can be measured at slightly higher energy, $\hbar \omega=1.45$~meV with integration in the same range. A point-by-point subtraction leaves the magnetic contribution plotted in Fig.~\ref{Fig-states}(c). There is clearly no modulation of intensity other than that along the chain-axis.

\begin{figure}
	\includegraphics[width=8.6 cm]{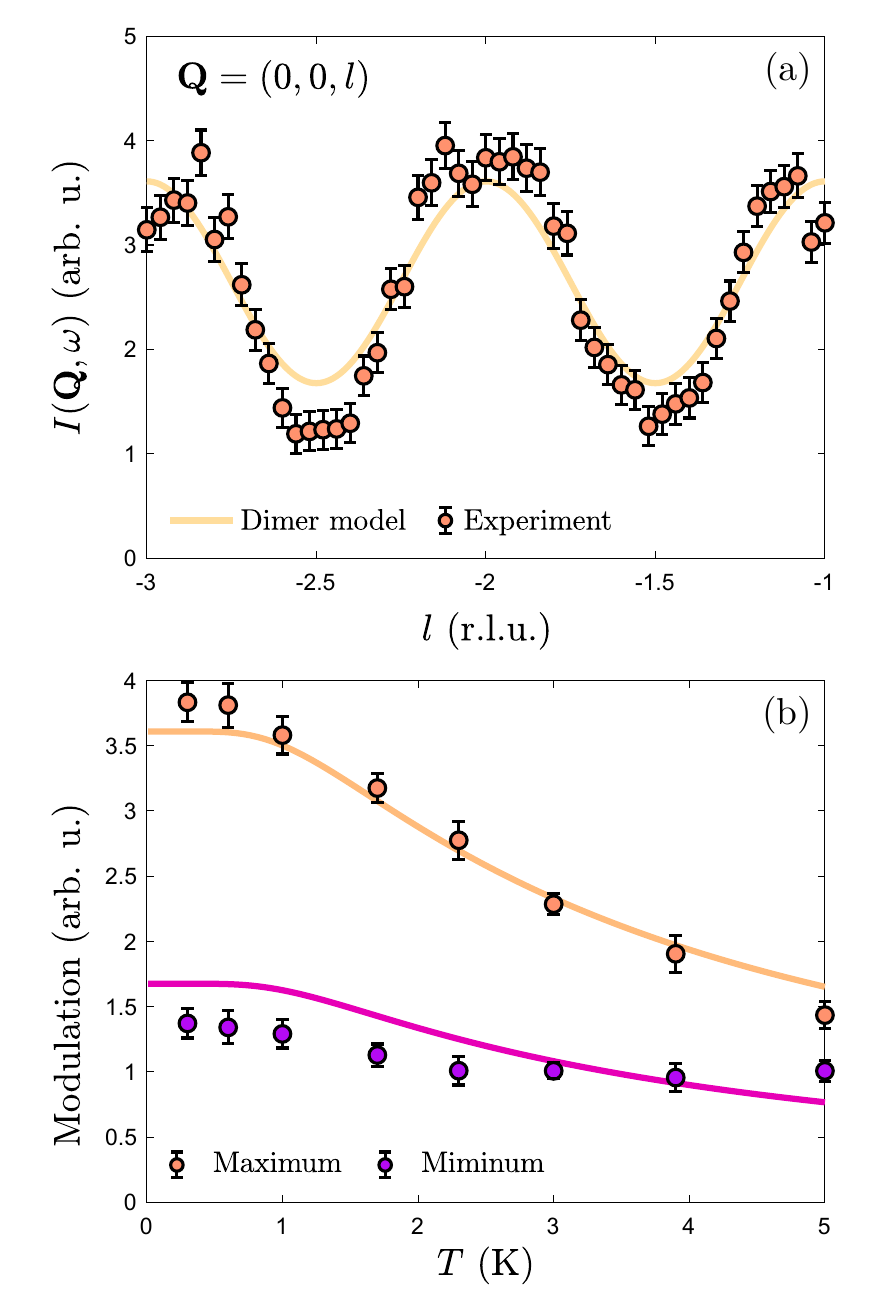}
	\caption{\label{Fig-dimer} Symbols: same data as in Fig.~\ref{Fig-states}(c), integrated over the $(h,h,0)$-direction. The solid line is a single-parameter dimer-model fit, as described in the text. (b) Temperature dependence of the intensity plotted in (a) for $l=-2$ (top) and $l=-2.5$ (bottom), respectively. The solid lines are calculations based on the dimer model.}
\end{figure}

The $c$-axis intensity modulation is quantified by integrating the data shown in Fig.~\ref{Fig-states}(c) in the range $\pm 0.75$~r.l.u. along the $(h,h,0)$ direction. It is plotted in solid symbols in Fig.~\ref{Fig-dimer}(a). The observed periodicity indeed corresponds to the spacing between two nearest-neighbor spins along the $c$ direction. The somewhat puzzling observation is that the intensity is maximal at {\em integer} values of $l$. For a Heisenberg $S=1$ dimer we expect the intensity to vanish at that point and the maximum to be located at half-integer $l$ instead. This mystery is solved if we consider the role of the strong easy-plane anisotropy in DTNX. With such anisotropy the total spin of the dimer is no longer a good quantum number and only its projection $S^z$ is conserved. The original triplet of $S=1$ excitation from the $S=0$ singlet ground state is split into a longitudinal and transverse components. These are still modulated so that intensity is a maximum at half-integer values of $l$ and vanishes at integer positions. However, the originally forbidden transition to the $S^z=\pm 1$ members of the $S=2$ quintuplet now becomes allowed and is transverse-polarized. The corresponding structure factor has a maximum at integer $l$. 

We consider the following toy model based on the Hamiltonian proposed in Ref.~\cite{DupontCapponiLaflorencie2017}:
\be
\hat{\mathcal{H}}=D'(\hat{S}_1^z)^2+D(\hat{S}_2^z)^2+J'\hat{\mathbf{S}}_1\hat{\mathbf{S}}_2.
\ee
Each Br impurity i) substantially increases the coupling constant $J'$ on the affected Ni--Br--Cl--Ni superexchange pathway compared to $J_c$ for unaffected Ni-Cl-Cl-Ni bonds, to the extent that the dimer can be considered as isolated (all other exchange interactions are disregarded), and ii) reduces the anisotropy constant $D'$ on the adjacent Ni$^{2+}$ spin, leaving the other one intact. Ref.~\cite{DupontCapponiLaflorencie2017} estimates $D'=0.28$~meV and $J'=0.46$~meV, respectively. Apart from these two largest parameters, other interactions are disregarded, resulting in isolated dimers. The model is easily diagonalized numerically. We calculate the two descendants of the triplet at $0.42$~meV and $1.01$~meV, respectively, and the $S^z=\pm 1$ quintuplet-descendants at $1.46$~meV. To match the latter to the observed energy of the impurity mode precisely, we somewhat arbitrarily select a slightly smaller value of the impurity-affected coupling constant, namely $J'=0.365$~meV. With this choice and only an overall scale factor as fitting parameter, the dimer model reproduces the observed intensity modulation remarkably well [solid line in Fig.~\ref{Fig-dimer}(a)]. The measured temperature dependence of the impurity-state intensity at the maximum and minimum, respectively, is plotted in Fig.~\ref{Fig-dimer}(b). It is also nicely reproduced by the dimer model without any additional adjustable parameters (solid lines). 

\begin{figure}
	\includegraphics[width=\columnwidth]{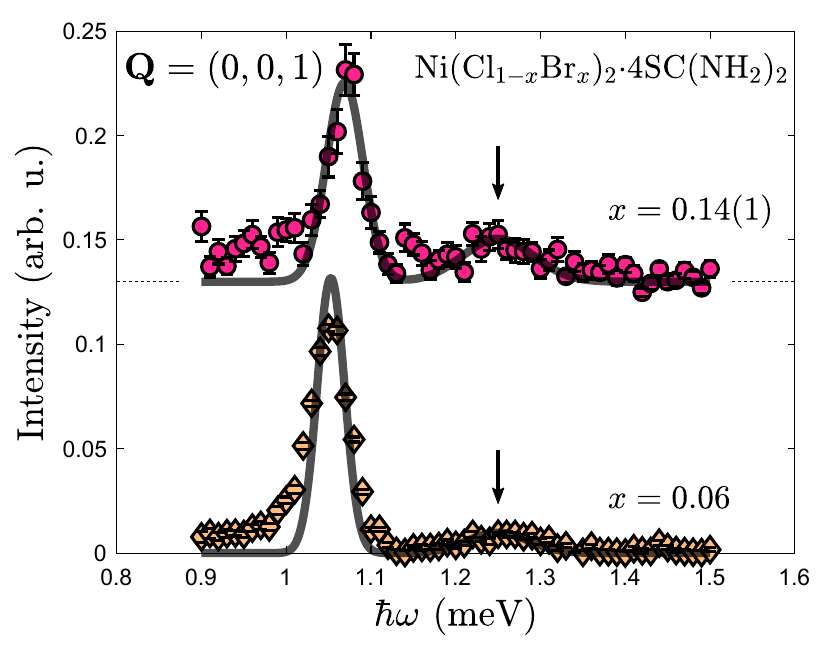}
	\caption{\label{comparison} Constant-energy scans across the magnon (lower peak) and local excitation (upper peak, indicated by the arrow) collected at the $(0,0,1)$ reciprocal-space point measured in DTNX at base temperature. The intensity is integrated in the range $\pm 0.1$~r.l.u. in the $(h,h,0)$, $(h,-h,0)$ and $\pm 0.05$~r.l.u. in the $(0,0,l)$ directions. Circles and diamonds correspond to nominal Br content of $x=0.14(1)$ and $x=0.06$. The latter data set is from the experiment described in Ref.~\cite{PovarovWulfHuvonen2015}. Solid lines are empyrical Gaussian fits. The $x=0.14$ data set is offset for visibility. }
\end{figure}

An additional consistency check comes from the analysis of the local mode's energy-integrated intensity. It can be properly normalized by comparing it to the intensity of the magnon branch, as illustrated in Fig.~\ref{comparison} (circles). Here we show an energy ``scan'' cut from the data at the $(0,0,1)$ position. The stronger and weaker peaks correspond to the magnon and local mode, respectively. We use empirical Gaussian fits to estimate their intensities (solid line). Comparing the measured intensity ratio to that obtained from the GSWT/RPA and direct diagonalization for the magnon and dimer model, respectively, we conclude that there are 0.36(2) dimer spins for every Ni$^{2+}$ ion involved in the magnon branch. If we assume that each Br center creates one dimer, this corresponds to a $x=0.18(1)$ Br-content, in very reasonable agreement with the nominal composition of our sample. The local mode intensity clearly scales with Br content. We performed a similar analysis of the data in Ref.~\cite{PovarovWulfHuvonen2015} for the nominally $x=0.06$ sample. The corresponding scan across the magnon and local mode is also plotted in Fig.~\ref{comparison} (diamonds). Estimating the intensities by Gaussian fits yields 0.16(2) dimer spins for every magnon spin, which corresponds to a Br content of 0.09(1), also in reasonable agreement with the nominal.

We can conclude that, at least as far as the 1.25~meV excitation is concerned, the simplistic single-strong-bond-dimer model works remarkably well. Of course, in other aspects it is over-simplified, since it does not consider the coupling of the strong-dimer to the rest of the Ni$^{2+}$ chain. The two descendants of the triplet mode in our model are calculated to be inside the magnon band, at 0.36~meV and 0.96~meV, respectively. There are, at best, only vague hints of flat bands at these energies in our present data, although such bands are better visible in the $x=0.06$ material (Fig.~7 of Ref.~\cite{PovarovWulfHuvonen2015}). Let us consider the intensities of these two calculated dimer excitations. The observation of the $c$-axis-polarized 0.96~meV mode would be suppressed by the polarization factor. For most of the data in Fig.~\ref{Fig-states} the angle between the scattering vector and $c$ axis is below $25^\circ$, which corresponds to a polarization factor of only 20\%, while the calculated structure factor of the 0.96~meV excitation is comparable to that of the observed 1.25~meV mode. Such an argument can't be made for the 0.36~meV excitation. It has the same polarization factor as the 1.25~meV transition but is calculated to be almost 5 times stronger. That intensity maximum is, however, located at half-integer $l$, where the excitation would overlap with a much stronger dispersive magnon band and may be difficult to observe.

The more important factor may be that the local dimer excitations lying {\em inside} the magnon band hybridize (scatter) with the latter. This is only natural, considering that in the material the dimer is certainly {\em not} isolated, but is incorporated into a chain, albeit one with weaker coupling constants. In both our $x=0.14$ sample and the previously studied $x=0.06$ material there is a clear excess of magnon intensity near the magnon dispersion saddle point $(0,0,0.5)$, as compared to the overall GSWT fit (see Fig.~\ref{Fig-disp}(b) above and Fig.~4(b) in Ref.~\cite{PovarovWulfHuvonen2015}). Here the magnon energy approximately matches that of the would-be 0.36~meV dimer mode. From our model we estimate the intensity of the latter to account for about a quarter of the intensity ``surplus'' at the saddle point.

\vspace{1cm}

\section{Conclusions}
In summary, to within experimental energy resolution, the spin excitations in DTNX near the critical Br-content do show finite-$T$ scaling as appropriate for a true quantum-critical point. The scaling exponent may, however, deviate from the Mean Field value expected in the absence of disorder.

The disorder-induced local excitations are to be associated with a specific transition in the strong-bond dimer that forms around each Br impurity, in agreement with the model of Ref.~\cite{DupontCapponiLaflorencie2017}. This interpretation accounts for the observed energies, intensities, wave vector and temperature dependencies of the local mode remarkably well.

\begin{acknowledgements}
This work was supported by Swiss National Science Foundation under Division II. AZ thanks Prof. Batista at U. Tennessee for helpful comments.
\end{acknowledgements}

\bibliographystyle{apsrev4-2}
\bibliography{v02_DTNX_main.bbl}

\begin{thebibliography}{43}%
\makeatletter
\providecommand \@ifxundefined [1]{%
 \@ifx{#1\undefined}
}%
\providecommand \@ifnum [1]{%
 \ifnum #1\expandafter \@firstoftwo
 \else \expandafter \@secondoftwo
 \fi
}%
\providecommand \@ifx [1]{%
 \ifx #1\expandafter \@firstoftwo
 \else \expandafter \@secondoftwo
 \fi
}%
\providecommand \natexlab [1]{#1}%
\providecommand \enquote  [1]{``#1''}%
\providecommand \bibnamefont  [1]{#1}%
\providecommand \bibfnamefont [1]{#1}%
\providecommand \citenamefont [1]{#1}%
\providecommand \href@noop [0]{\@secondoftwo}%
\providecommand \href [0]{\begingroup \@sanitize@url \@href}%
\providecommand \@href[1]{\@@startlink{#1}\@@href}%
\providecommand \@@href[1]{\endgroup#1\@@endlink}%
\providecommand \@sanitize@url [0]{\catcode `\\12\catcode `\$12\catcode
  `\&12\catcode `\#12\catcode `\^12\catcode `\_12\catcode `\%12\relax}%
\providecommand \@@startlink[1]{}%
\providecommand \@@endlink[0]{}%
\providecommand \url  [0]{\begingroup\@sanitize@url \@url }%
\providecommand \@url [1]{\endgroup\@href {#1}{\urlprefix }}%
\providecommand \urlprefix  [0]{URL }%
\providecommand \Eprint [0]{\href }%
\providecommand \doibase [0]{https://doi.org/}%
\providecommand \selectlanguage [0]{\@gobble}%
\providecommand \bibinfo  [0]{\@secondoftwo}%
\providecommand \bibfield  [0]{\@secondoftwo}%
\providecommand \translation [1]{[#1]}%
\providecommand \BibitemOpen [0]{}%
\providecommand \bibitemStop [0]{}%
\providecommand \bibitemNoStop [0]{.\EOS\space}%
\providecommand \EOS [0]{\spacefactor3000\relax}%
\providecommand \BibitemShut  [1]{\csname bibitem#1\endcsname}%
\let\auto@bib@innerbib\@empty
\bibitem [{\citenamefont {Giamarchi}\ \emph {et~al.}(2008)\citenamefont
  {Giamarchi}, \citenamefont {R\"uegg},\ and\ \citenamefont
  {Tchernyshyov}}]{GiamarchiRueggTchernyshyov2008}%
  \BibitemOpen
  \bibfield  {author} {\bibinfo {author} {\bibfnamefont {T.}~\bibnamefont
  {Giamarchi}}, \bibinfo {author} {\bibfnamefont {C.}~\bibnamefont {R\"uegg}},\
  and\ \bibinfo {author} {\bibfnamefont {O.}~\bibnamefont {Tchernyshyov}},\
  }\href {https://doi.org/10.1038/nphys893} {\bibfield  {journal} {\bibinfo
  {journal} {Nat. Phys.}\ }\textbf {\bibinfo {volume} {4}},\ \bibinfo {pages}
  {198} (\bibinfo {year} {2008})}\BibitemShut {NoStop}%
\bibitem [{\citenamefont {Zapf}\ \emph {et~al.}(2014)\citenamefont {Zapf},
  \citenamefont {Jaime},\ and\ \citenamefont {Batista}}]{ZapfJaimeBatista2014}%
  \BibitemOpen
  \bibfield  {author} {\bibinfo {author} {\bibfnamefont {V.}~\bibnamefont
  {Zapf}}, \bibinfo {author} {\bibfnamefont {M.}~\bibnamefont {Jaime}},\ and\
  \bibinfo {author} {\bibfnamefont {C.~D.}\ \bibnamefont {Batista}},\ }\href
  {https://doi.org/10.1103/RevModPhys.86.563} {\bibfield  {journal} {\bibinfo
  {journal} {Rev. Mod. Phys.}\ }\textbf {\bibinfo {volume} {86}},\ \bibinfo
  {pages} {563} (\bibinfo {year} {2014})}\BibitemShut {NoStop}%
\bibitem [{\citenamefont {Zheludev}(2020)}]{Zheludev2020}%
  \BibitemOpen
  \bibfield  {author} {\bibinfo {author} {\bibfnamefont {A.}~\bibnamefont
  {Zheludev}},\ }\href {https://doi.org/10.1134/S1063776120070183} {\bibfield
  {journal} {\bibinfo  {journal} {J. Exp. Theor. Phys.}\ }\textbf {\bibinfo
  {volume} {131}},\ \bibinfo {pages} {34} (\bibinfo {year} {2020})}\BibitemShut
  {NoStop}%
\bibitem [{\citenamefont {Tanaka}\ \emph {et~al.}(2003)\citenamefont {Tanaka},
  \citenamefont {Goto}, \citenamefont {Fujisawa}, \citenamefont {Ono},\ and\
  \citenamefont {Uwatoko}}]{TanakaGotoFujisawa2003}%
  \BibitemOpen
  \bibfield  {author} {\bibinfo {author} {\bibfnamefont {H.}~\bibnamefont
  {Tanaka}}, \bibinfo {author} {\bibfnamefont {K.}~\bibnamefont {Goto}},
  \bibinfo {author} {\bibfnamefont {M.}~\bibnamefont {Fujisawa}}, \bibinfo
  {author} {\bibfnamefont {T.}~\bibnamefont {Ono}},\ and\ \bibinfo {author}
  {\bibfnamefont {Y.}~\bibnamefont {Uwatoko}},\ }\href
  {https://doi.org/10.1016/S0921-4526(02)02009-4} {\bibfield  {journal}
  {\bibinfo  {journal} {Physica B}\ }\textbf {\bibinfo {volume} {329-333}},\
  \bibinfo {pages} {697} (\bibinfo {year} {2003})}\BibitemShut {NoStop}%
\bibitem [{\citenamefont {R\"uegg}\ \emph {et~al.}(2004)\citenamefont
  {R\"uegg}, \citenamefont {Furrer}, \citenamefont {Sheptyakov}, \citenamefont
  {Str\"assle}, \citenamefont {Kr\"amer}, \citenamefont {G\"udel},\ and\
  \citenamefont {M\'el\'esi}}]{RueggFurrerSheptyakov2004}%
  \BibitemOpen
  \bibfield  {author} {\bibinfo {author} {\bibfnamefont {C.}~\bibnamefont
  {R\"uegg}}, \bibinfo {author} {\bibfnamefont {A.}~\bibnamefont {Furrer}},
  \bibinfo {author} {\bibfnamefont {D.}~\bibnamefont {Sheptyakov}}, \bibinfo
  {author} {\bibfnamefont {T.}~\bibnamefont {Str\"assle}}, \bibinfo {author}
  {\bibfnamefont {K.~W.}\ \bibnamefont {Kr\"amer}}, \bibinfo {author}
  {\bibfnamefont {H.-U.}\ \bibnamefont {G\"udel}},\ and\ \bibinfo {author}
  {\bibfnamefont {L.}~\bibnamefont {M\'el\'esi}},\ }\href
  {https://doi.org/10.1103/PhysRevLett.93.257201} {\bibfield  {journal}
  {\bibinfo  {journal} {Phys. Rev. Lett.}\ }\textbf {\bibinfo {volume} {93}},\
  \bibinfo {pages} {257201} (\bibinfo {year} {2004})}\BibitemShut {NoStop}%
\bibitem [{\citenamefont {Thede}\ \emph {et~al.}(2014)\citenamefont {Thede},
  \citenamefont {Mannig}, \citenamefont {M\aa{}nsson}, \citenamefont
  {H\"uvonen}, \citenamefont {Khasanov}, \citenamefont {Morenzoni},\ and\
  \citenamefont {Zheludev}}]{ThedeMannigMansson2014}%
  \BibitemOpen
  \bibfield  {author} {\bibinfo {author} {\bibfnamefont {M.}~\bibnamefont
  {Thede}}, \bibinfo {author} {\bibfnamefont {A.}~\bibnamefont {Mannig}},
  \bibinfo {author} {\bibfnamefont {M.}~\bibnamefont {M\aa{}nsson}}, \bibinfo
  {author} {\bibfnamefont {D.}~\bibnamefont {H\"uvonen}}, \bibinfo {author}
  {\bibfnamefont {R.}~\bibnamefont {Khasanov}}, \bibinfo {author}
  {\bibfnamefont {E.}~\bibnamefont {Morenzoni}},\ and\ \bibinfo {author}
  {\bibfnamefont {A.}~\bibnamefont {Zheludev}},\ }\href
  {https://doi.org/10.1103/PhysRevLett.112.087204} {\bibfield  {journal}
  {\bibinfo  {journal} {Phys. Rev. Lett.}\ }\textbf {\bibinfo {volume} {112}},\
  \bibinfo {pages} {087204} (\bibinfo {year} {2014})}\BibitemShut {NoStop}%
\bibitem [{\citenamefont {Perren}\ \emph {et~al.}(2015)\citenamefont {Perren},
  \citenamefont {M\"oller}, \citenamefont {H\"uvonen}, \citenamefont
  {Podlesnyak},\ and\ \citenamefont {Zheludev}}]{PerrenMollerHuvonen2015}%
  \BibitemOpen
  \bibfield  {author} {\bibinfo {author} {\bibfnamefont {G.}~\bibnamefont
  {Perren}}, \bibinfo {author} {\bibfnamefont {J.~S.}\ \bibnamefont
  {M\"oller}}, \bibinfo {author} {\bibfnamefont {D.}~\bibnamefont {H\"uvonen}},
  \bibinfo {author} {\bibfnamefont {A.~A.}\ \bibnamefont {Podlesnyak}},\ and\
  \bibinfo {author} {\bibfnamefont {A.}~\bibnamefont {Zheludev}},\ }\href
  {https://doi.org/10.1103/PhysRevB.92.054413} {\bibfield  {journal} {\bibinfo
  {journal} {Phys. Rev. B}\ }\textbf {\bibinfo {volume} {92}},\ \bibinfo
  {pages} {054413} (\bibinfo {year} {2015})}\BibitemShut {NoStop}%
\bibitem [{\citenamefont {Kurita}\ and\ \citenamefont
  {Tanaka}(2016)}]{KuritaTanaka2016}%
  \BibitemOpen
  \bibfield  {author} {\bibinfo {author} {\bibfnamefont {N.}~\bibnamefont
  {Kurita}}\ and\ \bibinfo {author} {\bibfnamefont {H.}~\bibnamefont
  {Tanaka}},\ }\href {https://doi.org/10.1103/PhysRevB.94.104409} {\bibfield
  {journal} {\bibinfo  {journal} {Phys. Rev. B}\ }\textbf {\bibinfo {volume}
  {94}},\ \bibinfo {pages} {104409} (\bibinfo {year} {2016})}\BibitemShut
  {NoStop}%
\bibitem [{\citenamefont {Hayashida}\ \emph {et~al.}(2018)\citenamefont
  {Hayashida}, \citenamefont {Zaharko}, \citenamefont {Kurita}, \citenamefont
  {Tanaka}, \citenamefont {Hagihala}, \citenamefont {Soda}, \citenamefont
  {Itoh}, \citenamefont {Uwatoko},\ and\ \citenamefont
  {Masuda}}]{HayashidaZaharkoKurita2018}%
  \BibitemOpen
  \bibfield  {author} {\bibinfo {author} {\bibfnamefont {S.}~\bibnamefont
  {Hayashida}}, \bibinfo {author} {\bibfnamefont {O.}~\bibnamefont {Zaharko}},
  \bibinfo {author} {\bibfnamefont {N.}~\bibnamefont {Kurita}}, \bibinfo
  {author} {\bibfnamefont {H.}~\bibnamefont {Tanaka}}, \bibinfo {author}
  {\bibfnamefont {M.}~\bibnamefont {Hagihala}}, \bibinfo {author}
  {\bibfnamefont {M.}~\bibnamefont {Soda}}, \bibinfo {author} {\bibfnamefont
  {S.}~\bibnamefont {Itoh}}, \bibinfo {author} {\bibfnamefont {Y.}~\bibnamefont
  {Uwatoko}},\ and\ \bibinfo {author} {\bibfnamefont {T.}~\bibnamefont
  {Masuda}},\ }\href {https://doi.org/10.1103/PhysRevB.97.140405} {\bibfield
  {journal} {\bibinfo  {journal} {Phys. Rev. B}\ }\textbf {\bibinfo {volume}
  {97}},\ \bibinfo {pages} {140405(R)} (\bibinfo {year} {2018})}\BibitemShut
  {NoStop}%
\bibitem [{\citenamefont {Povarov}\ \emph {et~al.}(2017)\citenamefont
  {Povarov}, \citenamefont {Mannig}, \citenamefont {Perren}, \citenamefont
  {M\"oller}, \citenamefont {Wulf}, \citenamefont {Ollivier},\ and\
  \citenamefont {Zheludev}}]{PovarovMannigPerren2017}%
  \BibitemOpen
  \bibfield  {author} {\bibinfo {author} {\bibfnamefont {K.~Y.}\ \bibnamefont
  {Povarov}}, \bibinfo {author} {\bibfnamefont {A.}~\bibnamefont {Mannig}},
  \bibinfo {author} {\bibfnamefont {G.}~\bibnamefont {Perren}}, \bibinfo
  {author} {\bibfnamefont {J.~S.}\ \bibnamefont {M\"oller}}, \bibinfo {author}
  {\bibfnamefont {E.}~\bibnamefont {Wulf}}, \bibinfo {author} {\bibfnamefont
  {J.}~\bibnamefont {Ollivier}},\ and\ \bibinfo {author} {\bibfnamefont
  {A.}~\bibnamefont {Zheludev}},\ }\href
  {https://doi.org/10.1103/PhysRevB.96.140414} {\bibfield  {journal} {\bibinfo
  {journal} {Phys. Rev. B}\ }\textbf {\bibinfo {volume} {96}},\ \bibinfo
  {pages} {140414(R)} (\bibinfo {year} {2017})}\BibitemShut {NoStop}%
\bibitem [{\citenamefont {Hayashida}\ \emph {et~al.}(2019)\citenamefont
  {Hayashida}, \citenamefont {Stoppel}, \citenamefont {Yan}, \citenamefont
  {Gvasaliya}, \citenamefont {Podlesnyak},\ and\ \citenamefont
  {Zheludev}}]{HayashidaStoppelYan2019}%
  \BibitemOpen
  \bibfield  {author} {\bibinfo {author} {\bibfnamefont {S.}~\bibnamefont
  {Hayashida}}, \bibinfo {author} {\bibfnamefont {L.}~\bibnamefont {Stoppel}},
  \bibinfo {author} {\bibfnamefont {Z.}~\bibnamefont {Yan}}, \bibinfo {author}
  {\bibfnamefont {S.}~\bibnamefont {Gvasaliya}}, \bibinfo {author}
  {\bibfnamefont {A.}~\bibnamefont {Podlesnyak}},\ and\ \bibinfo {author}
  {\bibfnamefont {A.}~\bibnamefont {Zheludev}},\ }\href
  {https://doi.org/10.1103/PhysRevB.99.224420} {\bibfield  {journal} {\bibinfo
  {journal} {Phys. Rev. B}\ }\textbf {\bibinfo {volume} {99}},\ \bibinfo
  {pages} {224420} (\bibinfo {year} {2019})}\BibitemShut {NoStop}%
\bibitem [{\citenamefont {Yu}\ \emph {et~al.}(2012)\citenamefont {Yu},
  \citenamefont {Yin}, \citenamefont {Sullivan}, \citenamefont {Xia},
  \citenamefont {Huan}, \citenamefont {Paduan-Filho}, \citenamefont
  {Oliveira~Jr}, \citenamefont {Haas}, \citenamefont {Steppke}, \citenamefont
  {Miclea} \emph {et~al.}}]{YuYinSullivan2012}%
  \BibitemOpen
  \bibfield  {author} {\bibinfo {author} {\bibfnamefont {R.}~\bibnamefont
  {Yu}}, \bibinfo {author} {\bibfnamefont {L.}~\bibnamefont {Yin}}, \bibinfo
  {author} {\bibfnamefont {N.~S.}\ \bibnamefont {Sullivan}}, \bibinfo {author}
  {\bibfnamefont {J.}~\bibnamefont {Xia}}, \bibinfo {author} {\bibfnamefont
  {C.}~\bibnamefont {Huan}}, \bibinfo {author} {\bibfnamefont {A.}~\bibnamefont
  {Paduan-Filho}}, \bibinfo {author} {\bibfnamefont {N.~F.}\ \bibnamefont
  {Oliveira~Jr}}, \bibinfo {author} {\bibfnamefont {S.}~\bibnamefont {Haas}},
  \bibinfo {author} {\bibfnamefont {A.}~\bibnamefont {Steppke}}, \bibinfo
  {author} {\bibfnamefont {C.~F.}\ \bibnamefont {Miclea}}, \emph {et~al.},\
  }\href {https://doi.org/10.1038/nature11406} {\bibfield  {journal} {\bibinfo
  {journal} {Nature (London)}\ }\textbf {\bibinfo {volume} {489}},\ \bibinfo
  {pages} {379} (\bibinfo {year} {2012})}\BibitemShut {NoStop}%
\bibitem [{\citenamefont {Paduan‐Filho}\ \emph {et~al.}(1981)\citenamefont
  {Paduan‐Filho}, \citenamefont {Chirico}, \citenamefont {Joung},\ and\
  \citenamefont {Carlin}}]{PaduanFilhoChiricoJoung1981}%
  \BibitemOpen
  \bibfield  {author} {\bibinfo {author} {\bibfnamefont {A.}~\bibnamefont
  {Paduan‐Filho}}, \bibinfo {author} {\bibfnamefont {R.~D.}\ \bibnamefont
  {Chirico}}, \bibinfo {author} {\bibfnamefont {K.~O.}\ \bibnamefont {Joung}},\
  and\ \bibinfo {author} {\bibfnamefont {R.~L.}\ \bibnamefont {Carlin}},\
  }\href {https://doi.org/10.1063/1.441589} {\bibfield  {journal} {\bibinfo
  {journal} {J. Chem. Phys.}\ }\textbf {\bibinfo {volume} {74}},\ \bibinfo
  {pages} {4103} (\bibinfo {year} {1981})}\BibitemShut {NoStop}%
\bibitem [{\citenamefont {Zapf}\ \emph {et~al.}(2006)\citenamefont {Zapf},
  \citenamefont {Zocco}, \citenamefont {Hansen}, \citenamefont {Jaime},
  \citenamefont {Harrison}, \citenamefont {Batista}, \citenamefont
  {Kenzelmann}, \citenamefont {Niedermayer}, \citenamefont {Lacerda},\ and\
  \citenamefont {Paduan-Filho}}]{ZapfZoccoHansen2006}%
  \BibitemOpen
  \bibfield  {author} {\bibinfo {author} {\bibfnamefont {V.~S.}\ \bibnamefont
  {Zapf}}, \bibinfo {author} {\bibfnamefont {D.}~\bibnamefont {Zocco}},
  \bibinfo {author} {\bibfnamefont {B.~R.}\ \bibnamefont {Hansen}}, \bibinfo
  {author} {\bibfnamefont {M.}~\bibnamefont {Jaime}}, \bibinfo {author}
  {\bibfnamefont {N.}~\bibnamefont {Harrison}}, \bibinfo {author}
  {\bibfnamefont {C.~D.}\ \bibnamefont {Batista}}, \bibinfo {author}
  {\bibfnamefont {M.}~\bibnamefont {Kenzelmann}}, \bibinfo {author}
  {\bibfnamefont {C.}~\bibnamefont {Niedermayer}}, \bibinfo {author}
  {\bibfnamefont {A.}~\bibnamefont {Lacerda}},\ and\ \bibinfo {author}
  {\bibfnamefont {A.}~\bibnamefont {Paduan-Filho}},\ }\href
  {https://doi.org/10.1103/PhysRevLett.96.077204} {\bibfield  {journal}
  {\bibinfo  {journal} {Phys. Rev. Lett.}\ }\textbf {\bibinfo {volume} {96}},\
  \bibinfo {pages} {077204} (\bibinfo {year} {2006})}\BibitemShut {NoStop}%
\bibitem [{\citenamefont {Zvyagin}\ \emph {et~al.}(2007)\citenamefont
  {Zvyagin}, \citenamefont {Wosnitza}, \citenamefont {Batista}, \citenamefont
  {Tsukamoto}, \citenamefont {Kawashima}, \citenamefont {Krzystek},
  \citenamefont {Zapf}, \citenamefont {Jaime}, \citenamefont {Oliveira},\ and\
  \citenamefont {Paduan-Filho}}]{ZvyaginWosnitzaBatista2007}%
  \BibitemOpen
  \bibfield  {author} {\bibinfo {author} {\bibfnamefont {S.~A.}\ \bibnamefont
  {Zvyagin}}, \bibinfo {author} {\bibfnamefont {J.}~\bibnamefont {Wosnitza}},
  \bibinfo {author} {\bibfnamefont {C.~D.}\ \bibnamefont {Batista}}, \bibinfo
  {author} {\bibfnamefont {M.}~\bibnamefont {Tsukamoto}}, \bibinfo {author}
  {\bibfnamefont {N.}~\bibnamefont {Kawashima}}, \bibinfo {author}
  {\bibfnamefont {J.}~\bibnamefont {Krzystek}}, \bibinfo {author}
  {\bibfnamefont {V.~S.}\ \bibnamefont {Zapf}}, \bibinfo {author}
  {\bibfnamefont {M.}~\bibnamefont {Jaime}}, \bibinfo {author} {\bibfnamefont
  {N.~F.}\ \bibnamefont {Oliveira}},\ and\ \bibinfo {author} {\bibfnamefont
  {A.}~\bibnamefont {Paduan-Filho}},\ }\href
  {https://doi.org/10.1103/PhysRevLett.98.047205} {\bibfield  {journal}
  {\bibinfo  {journal} {Phys. Rev. Lett.}\ }\textbf {\bibinfo {volume} {98}},\
  \bibinfo {pages} {047205} (\bibinfo {year} {2007})}\BibitemShut {NoStop}%
\bibitem [{\citenamefont {Yin}\ \emph {et~al.}(2008)\citenamefont {Yin},
  \citenamefont {Xia}, \citenamefont {Zapf}, \citenamefont {Sullivan},\ and\
  \citenamefont {Paduan-Filho}}]{YinXiaZapf2008}%
  \BibitemOpen
  \bibfield  {author} {\bibinfo {author} {\bibfnamefont {L.}~\bibnamefont
  {Yin}}, \bibinfo {author} {\bibfnamefont {J.~S.}\ \bibnamefont {Xia}},
  \bibinfo {author} {\bibfnamefont {V.~S.}\ \bibnamefont {Zapf}}, \bibinfo
  {author} {\bibfnamefont {N.~S.}\ \bibnamefont {Sullivan}},\ and\ \bibinfo
  {author} {\bibfnamefont {A.}~\bibnamefont {Paduan-Filho}},\ }\href
  {https://doi.org/10.1103/PhysRevLett.101.187205} {\bibfield  {journal}
  {\bibinfo  {journal} {Phys. Rev. Lett.}\ }\textbf {\bibinfo {volume} {101}},\
  \bibinfo {pages} {187205} (\bibinfo {year} {2008})}\BibitemShut {NoStop}%
\bibitem [{\citenamefont {Mukhopadhyay}\ \emph {et~al.}(2012)\citenamefont
  {Mukhopadhyay}, \citenamefont {Klanj\ifmmode~\check{s}\else \v{s}\fi{}ek},
  \citenamefont {Grbi\ifmmode~\acute{c}\else \'{c}\fi{}}, \citenamefont
  {Blinder}, \citenamefont {Mayaffre}, \citenamefont {Berthier}, \citenamefont
  {Horvati\ifmmode~\acute{c}\else \'{c}\fi{}}, \citenamefont {Continentino},
  \citenamefont {Paduan-Filho}, \citenamefont {Chiari},\ and\ \citenamefont
  {Piovesana}}]{MukhopadhyayKlanjsekGrbic2012}%
  \BibitemOpen
  \bibfield  {author} {\bibinfo {author} {\bibfnamefont {S.}~\bibnamefont
  {Mukhopadhyay}}, \bibinfo {author} {\bibfnamefont {M.}~\bibnamefont
  {Klanj\ifmmode~\check{s}\else \v{s}\fi{}ek}}, \bibinfo {author}
  {\bibfnamefont {M.~S.}\ \bibnamefont {Grbi\ifmmode~\acute{c}\else
  \'{c}\fi{}}}, \bibinfo {author} {\bibfnamefont {R.}~\bibnamefont {Blinder}},
  \bibinfo {author} {\bibfnamefont {H.}~\bibnamefont {Mayaffre}}, \bibinfo
  {author} {\bibfnamefont {C.}~\bibnamefont {Berthier}}, \bibinfo {author}
  {\bibfnamefont {M.}~\bibnamefont {Horvati\ifmmode~\acute{c}\else
  \'{c}\fi{}}}, \bibinfo {author} {\bibfnamefont {M.~A.}\ \bibnamefont
  {Continentino}}, \bibinfo {author} {\bibfnamefont {A.}~\bibnamefont
  {Paduan-Filho}}, \bibinfo {author} {\bibfnamefont {B.}~\bibnamefont
  {Chiari}},\ and\ \bibinfo {author} {\bibfnamefont {O.}~\bibnamefont
  {Piovesana}},\ }\href {https://doi.org/10.1103/PhysRevLett.109.177206}
  {\bibfield  {journal} {\bibinfo  {journal} {Phys. Rev. Lett.}\ }\textbf
  {\bibinfo {volume} {109}},\ \bibinfo {pages} {177206} (\bibinfo {year}
  {2012})}\BibitemShut {NoStop}%
\bibitem [{\citenamefont {Tsyrulin}\ \emph {et~al.}(2013)\citenamefont
  {Tsyrulin}, \citenamefont {Batista}, \citenamefont {Zapf}, \citenamefont
  {Jaime}, \citenamefont {Hansen}, \citenamefont {Niedermayer}, \citenamefont
  {Rule}, \citenamefont {Habicht}, \citenamefont {Prokes}, \citenamefont
  {Kiefer}, \citenamefont {Ressouche}, \citenamefont {Paduan-Filho},\ and\
  \citenamefont {Kenzelmann}}]{TsyrulinBatistaZapf2013}%
  \BibitemOpen
  \bibfield  {author} {\bibinfo {author} {\bibfnamefont {N.}~\bibnamefont
  {Tsyrulin}}, \bibinfo {author} {\bibfnamefont {C.~D.}\ \bibnamefont
  {Batista}}, \bibinfo {author} {\bibfnamefont {V.~S.}\ \bibnamefont {Zapf}},
  \bibinfo {author} {\bibfnamefont {M.}~\bibnamefont {Jaime}}, \bibinfo
  {author} {\bibfnamefont {B.~R.}\ \bibnamefont {Hansen}}, \bibinfo {author}
  {\bibfnamefont {C.}~\bibnamefont {Niedermayer}}, \bibinfo {author}
  {\bibfnamefont {K.~C.}\ \bibnamefont {Rule}}, \bibinfo {author}
  {\bibfnamefont {K.}~\bibnamefont {Habicht}}, \bibinfo {author} {\bibfnamefont
  {K.}~\bibnamefont {Prokes}}, \bibinfo {author} {\bibfnamefont
  {K.}~\bibnamefont {Kiefer}}, \bibinfo {author} {\bibfnamefont
  {E.}~\bibnamefont {Ressouche}}, \bibinfo {author} {\bibfnamefont
  {A.}~\bibnamefont {Paduan-Filho}},\ and\ \bibinfo {author} {\bibfnamefont
  {M.}~\bibnamefont {Kenzelmann}},\ }\href
  {https://doi.org/10.1088/0953-8984/25/21/216008} {\bibfield  {journal}
  {\bibinfo  {journal} {J. Phys. Condens. Matter}\ }\textbf {\bibinfo {volume}
  {25}},\ \bibinfo {pages} {216008} (\bibinfo {year} {2013})}\BibitemShut
  {NoStop}%
\bibitem [{\citenamefont {Wulf}\ \emph {et~al.}(2015)\citenamefont {Wulf},
  \citenamefont {H\"uvonen}, \citenamefont {Sch\"onemann}, \citenamefont
  {K\"uhne}, \citenamefont {Herrmannsd\"orfer}, \citenamefont {Glavatskyy},
  \citenamefont {Gerischer}, \citenamefont {Kiefer}, \citenamefont
  {Gvasaliya},\ and\ \citenamefont {Zheludev}}]{WulfHuvonenSchonemann2015}%
  \BibitemOpen
  \bibfield  {author} {\bibinfo {author} {\bibfnamefont {E.}~\bibnamefont
  {Wulf}}, \bibinfo {author} {\bibfnamefont {D.}~\bibnamefont {H\"uvonen}},
  \bibinfo {author} {\bibfnamefont {R.}~\bibnamefont {Sch\"onemann}}, \bibinfo
  {author} {\bibfnamefont {H.}~\bibnamefont {K\"uhne}}, \bibinfo {author}
  {\bibfnamefont {T.}~\bibnamefont {Herrmannsd\"orfer}}, \bibinfo {author}
  {\bibfnamefont {I.}~\bibnamefont {Glavatskyy}}, \bibinfo {author}
  {\bibfnamefont {S.}~\bibnamefont {Gerischer}}, \bibinfo {author}
  {\bibfnamefont {K.}~\bibnamefont {Kiefer}}, \bibinfo {author} {\bibfnamefont
  {S.}~\bibnamefont {Gvasaliya}},\ and\ \bibinfo {author} {\bibfnamefont
  {A.}~\bibnamefont {Zheludev}},\ }\href
  {https://doi.org/10.1103/PhysRevB.91.014406} {\bibfield  {journal} {\bibinfo
  {journal} {Phys. Rev. B}\ }\textbf {\bibinfo {volume} {91}},\ \bibinfo
  {pages} {014406} (\bibinfo {year} {2015})}\BibitemShut {NoStop}%
\bibitem [{\citenamefont {Povarov}\ \emph
  {et~al.}(2015{\natexlab{a}})\citenamefont {Povarov}, \citenamefont {Wulf},
  \citenamefont {H\"uvonen}, \citenamefont {Ollivier}, \citenamefont
  {Paduan-Filho},\ and\ \citenamefont {Zheludev}}]{PovarovWulfHuvonen2015}%
  \BibitemOpen
  \bibfield  {author} {\bibinfo {author} {\bibfnamefont {K.~Y.}\ \bibnamefont
  {Povarov}}, \bibinfo {author} {\bibfnamefont {E.}~\bibnamefont {Wulf}},
  \bibinfo {author} {\bibfnamefont {D.}~\bibnamefont {H\"uvonen}}, \bibinfo
  {author} {\bibfnamefont {J.}~\bibnamefont {Ollivier}}, \bibinfo {author}
  {\bibfnamefont {A.}~\bibnamefont {Paduan-Filho}},\ and\ \bibinfo {author}
  {\bibfnamefont {A.}~\bibnamefont {Zheludev}},\ }\href
  {https://doi.org/10.1103/PhysRevB.92.024429} {\bibfield  {journal} {\bibinfo
  {journal} {Phys. Rev. B}\ }\textbf {\bibinfo {volume} {92}},\ \bibinfo
  {pages} {024429} (\bibinfo {year} {2015}{\natexlab{a}})}\BibitemShut
  {NoStop}%
\bibitem [{\citenamefont {Mannig}\ \emph {et~al.}(2018)\citenamefont {Mannig},
  \citenamefont {Povarov}, \citenamefont {Ollivier},\ and\ \citenamefont
  {Zheludev}}]{MannigPovarovOllivier2018}%
  \BibitemOpen
  \bibfield  {author} {\bibinfo {author} {\bibfnamefont {A.}~\bibnamefont
  {Mannig}}, \bibinfo {author} {\bibfnamefont {K.~Y.}\ \bibnamefont {Povarov}},
  \bibinfo {author} {\bibfnamefont {J.}~\bibnamefont {Ollivier}},\ and\
  \bibinfo {author} {\bibfnamefont {A.}~\bibnamefont {Zheludev}},\ }\href
  {https://doi.org/10.1103/PhysRevB.98.214419} {\bibfield  {journal} {\bibinfo
  {journal} {Phys. Rev. B}\ }\textbf {\bibinfo {volume} {98}},\ \bibinfo
  {pages} {214419} (\bibinfo {year} {2018})}\BibitemShut {NoStop}%
\bibitem [{\citenamefont {Harris}(1974)}]{Harris1974}%
  \BibitemOpen
  \bibfield  {author} {\bibinfo {author} {\bibfnamefont {A.~B.}\ \bibnamefont
  {Harris}},\ }\href {https://doi.org/10.1088/0022-3719/7/9/009} {\bibfield
  {journal} {\bibinfo  {journal} {J. Phys. C}\ }\textbf {\bibinfo {volume}
  {7}},\ \bibinfo {pages} {1671} (\bibinfo {year} {1974})}\BibitemShut
  {NoStop}%
\bibitem [{\citenamefont {Vojta}(2013)}]{Vojta2013}%
  \BibitemOpen
  \bibfield  {author} {\bibinfo {author} {\bibfnamefont {M.}~\bibnamefont
  {Vojta}},\ }\href {https://doi.org/10.1103/PhysRevLett.111.097202} {\bibfield
   {journal} {\bibinfo  {journal} {Phys. Rev. Lett.}\ }\textbf {\bibinfo
  {volume} {111}},\ \bibinfo {pages} {097202} (\bibinfo {year}
  {2013})}\BibitemShut {NoStop}%
\bibitem [{\citenamefont {Paduan-Filho}\ \emph {et~al.}(2004)\citenamefont
  {Paduan-Filho}, \citenamefont {Gratens},\ and\ \citenamefont
  {Oliveira}}]{PaduanFilhoGratensOliveira2004}%
  \BibitemOpen
  \bibfield  {author} {\bibinfo {author} {\bibfnamefont {A.}~\bibnamefont
  {Paduan-Filho}}, \bibinfo {author} {\bibfnamefont {X.}~\bibnamefont
  {Gratens}},\ and\ \bibinfo {author} {\bibfnamefont {N.~F.}\ \bibnamefont
  {Oliveira}},\ }\href {https://doi.org/10.1103/PhysRevB.69.020405} {\bibfield
  {journal} {\bibinfo  {journal} {Phys. Rev. B}\ }\textbf {\bibinfo {volume}
  {69}},\ \bibinfo {pages} {020405(R)} (\bibinfo {year} {2004})}\BibitemShut
  {NoStop}%
\bibitem [{\citenamefont {H\"uvonen}\ \emph {et~al.}(2012)\citenamefont
  {H\"uvonen}, \citenamefont {Zhao}, \citenamefont {M\aa{}nsson}, \citenamefont
  {Yankova}, \citenamefont {Ressouche}, \citenamefont {Niedermayer},
  \citenamefont {Laver}, \citenamefont {Gvasaliya},\ and\ \citenamefont
  {Zheludev}}]{HuvonenZhaoMansson2012}%
  \BibitemOpen
  \bibfield  {author} {\bibinfo {author} {\bibfnamefont {D.}~\bibnamefont
  {H\"uvonen}}, \bibinfo {author} {\bibfnamefont {S.}~\bibnamefont {Zhao}},
  \bibinfo {author} {\bibfnamefont {M.}~\bibnamefont {M\aa{}nsson}}, \bibinfo
  {author} {\bibfnamefont {T.}~\bibnamefont {Yankova}}, \bibinfo {author}
  {\bibfnamefont {E.}~\bibnamefont {Ressouche}}, \bibinfo {author}
  {\bibfnamefont {C.}~\bibnamefont {Niedermayer}}, \bibinfo {author}
  {\bibfnamefont {M.}~\bibnamefont {Laver}}, \bibinfo {author} {\bibfnamefont
  {S.~N.}\ \bibnamefont {Gvasaliya}},\ and\ \bibinfo {author} {\bibfnamefont
  {A.}~\bibnamefont {Zheludev}},\ }\href
  {https://doi.org/10.1103/PhysRevB.85.100410} {\bibfield  {journal} {\bibinfo
  {journal} {Phys. Rev. B}\ }\textbf {\bibinfo {volume} {85}},\ \bibinfo
  {pages} {100410(R)} (\bibinfo {year} {2012})}\BibitemShut {NoStop}%
\bibitem [{\citenamefont {Kohama}\ \emph {et~al.}(2011)\citenamefont {Kohama},
  \citenamefont {Sologubenko}, \citenamefont {Dilley}, \citenamefont {Zapf},
  \citenamefont {Jaime}, \citenamefont {Mydosh}, \citenamefont {Paduan-Filho},
  \citenamefont {Al-Hassanieh}, \citenamefont {Sengupta}, \citenamefont
  {Gangadharaiah}, \citenamefont {Chernyshev},\ and\ \citenamefont
  {Batista}}]{KohamaSologubenkoDilley2011}%
  \BibitemOpen
  \bibfield  {author} {\bibinfo {author} {\bibfnamefont {Y.}~\bibnamefont
  {Kohama}}, \bibinfo {author} {\bibfnamefont {A.~V.}\ \bibnamefont
  {Sologubenko}}, \bibinfo {author} {\bibfnamefont {N.~R.}\ \bibnamefont
  {Dilley}}, \bibinfo {author} {\bibfnamefont {V.~S.}\ \bibnamefont {Zapf}},
  \bibinfo {author} {\bibfnamefont {M.}~\bibnamefont {Jaime}}, \bibinfo
  {author} {\bibfnamefont {J.~A.}\ \bibnamefont {Mydosh}}, \bibinfo {author}
  {\bibfnamefont {A.}~\bibnamefont {Paduan-Filho}}, \bibinfo {author}
  {\bibfnamefont {K.~A.}\ \bibnamefont {Al-Hassanieh}}, \bibinfo {author}
  {\bibfnamefont {P.}~\bibnamefont {Sengupta}}, \bibinfo {author}
  {\bibfnamefont {S.}~\bibnamefont {Gangadharaiah}}, \bibinfo {author}
  {\bibfnamefont {A.~L.}\ \bibnamefont {Chernyshev}},\ and\ \bibinfo {author}
  {\bibfnamefont {C.~D.}\ \bibnamefont {Batista}},\ }\href
  {https://doi.org/10.1103/PhysRevLett.106.037203} {\bibfield  {journal}
  {\bibinfo  {journal} {Phys. Rev. Lett.}\ }\textbf {\bibinfo {volume} {106}},\
  \bibinfo {pages} {037203} (\bibinfo {year} {2011})}\BibitemShut {NoStop}%
\bibitem [{\citenamefont {Wulf}\ \emph {et~al.}(2011)\citenamefont {Wulf},
  \citenamefont {M\"uhlbauer}, \citenamefont {Yankova},\ and\ \citenamefont
  {Zheludev}}]{WulfMuhlbauerYankova2011}%
  \BibitemOpen
  \bibfield  {author} {\bibinfo {author} {\bibfnamefont {E.}~\bibnamefont
  {Wulf}}, \bibinfo {author} {\bibfnamefont {S.}~\bibnamefont {M\"uhlbauer}},
  \bibinfo {author} {\bibfnamefont {T.}~\bibnamefont {Yankova}},\ and\ \bibinfo
  {author} {\bibfnamefont {A.}~\bibnamefont {Zheludev}},\ }\href
  {https://doi.org/10.1103/PhysRevB.84.174414} {\bibfield  {journal} {\bibinfo
  {journal} {Phys. Rev. B}\ }\textbf {\bibinfo {volume} {84}},\ \bibinfo
  {pages} {174414} (\bibinfo {year} {2011})}\BibitemShut {NoStop}%
\bibitem [{\citenamefont {Squires}(1978)}]{Squiresbook1978}%
  \BibitemOpen
  \bibfield  {author} {\bibinfo {author} {\bibfnamefont {G.~L.}\ \bibnamefont
  {Squires}},\ }\href {https://doi.org/10.1017/CBO9781139107808} {\emph
  {\bibinfo {title} {{Introduction To The theory Of Thermal neutron
  Scattering}}}}\ (\bibinfo  {publisher} {Cambridge University Press,
  Cambridge, UK},\ \bibinfo {year} {1978})\BibitemShut {NoStop}%
\bibitem [{\citenamefont {Matsumoto}\ and\ \citenamefont
  {Koga}(2007)}]{MatsumotoKoga2007}%
  \BibitemOpen
  \bibfield  {author} {\bibinfo {author} {\bibfnamefont {M.}~\bibnamefont
  {Matsumoto}}\ and\ \bibinfo {author} {\bibfnamefont {M.}~\bibnamefont
  {Koga}},\ }\href {https://doi.org/10.1143/JPSJ.76.073709} {\bibfield
  {journal} {\bibinfo  {journal} {J. Phys. Soc. Jpn.}\ }\textbf {\bibinfo
  {volume} {76}},\ \bibinfo {pages} {073709} (\bibinfo {year}
  {2007})}\BibitemShut {NoStop}%
\bibitem [{\citenamefont {Zhang}\ \emph {et~al.}(2013)\citenamefont {Zhang},
  \citenamefont {Wierschem}, \citenamefont {Yap}, \citenamefont {Kato},
  \citenamefont {Batista},\ and\ \citenamefont
  {Sengupta}}]{ZhangWierschemYap2013}%
  \BibitemOpen
  \bibfield  {author} {\bibinfo {author} {\bibfnamefont {Z.}~\bibnamefont
  {Zhang}}, \bibinfo {author} {\bibfnamefont {K.}~\bibnamefont {Wierschem}},
  \bibinfo {author} {\bibfnamefont {I.}~\bibnamefont {Yap}}, \bibinfo {author}
  {\bibfnamefont {Y.}~\bibnamefont {Kato}}, \bibinfo {author} {\bibfnamefont
  {C.~D.}\ \bibnamefont {Batista}},\ and\ \bibinfo {author} {\bibfnamefont
  {P.}~\bibnamefont {Sengupta}},\ }\href
  {https://doi.org/10.1103/PhysRevB.87.174405} {\bibfield  {journal} {\bibinfo
  {journal} {Phys. Rev. B}\ }\textbf {\bibinfo {volume} {87}},\ \bibinfo
  {pages} {174405} (\bibinfo {year} {2013})}\BibitemShut {NoStop}%
\bibitem [{\citenamefont {Muniz}\ \emph {et~al.}(2014)\citenamefont {Muniz},
  \citenamefont {Kato},\ and\ \citenamefont {Batista}}]{MunizKatoYasuyuki2014}%
  \BibitemOpen
  \bibfield  {author} {\bibinfo {author} {\bibfnamefont {R.~A.}\ \bibnamefont
  {Muniz}}, \bibinfo {author} {\bibfnamefont {Y.}~\bibnamefont {Kato}},\ and\
  \bibinfo {author} {\bibfnamefont {C.~D.}\ \bibnamefont {Batista}},\ }\href
  {https://doi.org/10.1093/ptep/ptu109} {\bibfield  {journal} {\bibinfo
  {journal} {Prog. Theor. Exp. Phys.}\ }\textbf {\bibinfo {volume} {2014}},\
  \bibinfo {pages} {083I01} (\bibinfo {year} {2014})}\BibitemShut {NoStop}%
\bibitem [{\citenamefont {Sachdev}(1999)}]{SachdevBook2011}%
  \BibitemOpen
  \bibfield  {author} {\bibinfo {author} {\bibfnamefont {S.}~\bibnamefont
  {Sachdev}},\ }\href {https://doi.org/10.1017/CBO9780511973765} {\emph
  {\bibinfo {title} {{Quantum Phase Transitions}}}}\ (\bibinfo  {publisher}
  {Cambridge University Press, Cambridge, UK},\ \bibinfo {year}
  {1999})\BibitemShut {NoStop}%
\bibitem [{\citenamefont {Sachdev}\ and\ \citenamefont
  {Keimer}(2011)}]{SachdevKeimer2011}%
  \BibitemOpen
  \bibfield  {author} {\bibinfo {author} {\bibfnamefont {S.}~\bibnamefont
  {Sachdev}}\ and\ \bibinfo {author} {\bibfnamefont {B.}~\bibnamefont
  {Keimer}},\ }\href {https://doi.org/10.1063/1.3554314} {\bibfield  {journal}
  {\bibinfo  {journal} {Phys. Today}\ }\textbf {\bibinfo {volume} {64}},\
  \bibinfo {pages} {29} (\bibinfo {year} {2011})}\BibitemShut {NoStop}%
\bibitem [{\citenamefont {Blosser}\ \emph {et~al.}(2017)\citenamefont
  {Blosser}, \citenamefont {Kestin}, \citenamefont {Povarov}, \citenamefont
  {Bewley}, \citenamefont {Coira}, \citenamefont {Giamarchi},\ and\
  \citenamefont {Zheludev}}]{BlosserKestinPovarov2017}%
  \BibitemOpen
  \bibfield  {author} {\bibinfo {author} {\bibfnamefont {D.}~\bibnamefont
  {Blosser}}, \bibinfo {author} {\bibfnamefont {N.}~\bibnamefont {Kestin}},
  \bibinfo {author} {\bibfnamefont {K.~Y.}\ \bibnamefont {Povarov}}, \bibinfo
  {author} {\bibfnamefont {R.}~\bibnamefont {Bewley}}, \bibinfo {author}
  {\bibfnamefont {E.}~\bibnamefont {Coira}}, \bibinfo {author} {\bibfnamefont
  {T.}~\bibnamefont {Giamarchi}},\ and\ \bibinfo {author} {\bibfnamefont
  {A.}~\bibnamefont {Zheludev}},\ }\href
  {https://doi.org/10.1103/PhysRevB.96.134406} {\bibfield  {journal} {\bibinfo
  {journal} {Phys. Rev. B}\ }\textbf {\bibinfo {volume} {96}},\ \bibinfo
  {pages} {134406} (\bibinfo {year} {2017})}\BibitemShut {NoStop}%
\bibitem [{\citenamefont {Blosser}\ \emph {et~al.}(2018)\citenamefont
  {Blosser}, \citenamefont {Bhartiya}, \citenamefont {Voneshen},\ and\
  \citenamefont {Zheludev}}]{BlosserBhartiyaVoneshen2018}%
  \BibitemOpen
  \bibfield  {author} {\bibinfo {author} {\bibfnamefont {D.}~\bibnamefont
  {Blosser}}, \bibinfo {author} {\bibfnamefont {V.~K.}\ \bibnamefont
  {Bhartiya}}, \bibinfo {author} {\bibfnamefont {D.~J.}\ \bibnamefont
  {Voneshen}},\ and\ \bibinfo {author} {\bibfnamefont {A.}~\bibnamefont
  {Zheludev}},\ }\href {https://doi.org/10.1103/PhysRevLett.121.247201}
  {\bibfield  {journal} {\bibinfo  {journal} {Phys. Rev. Lett.}\ }\textbf
  {\bibinfo {volume} {121}},\ \bibinfo {pages} {247201} (\bibinfo {year}
  {2018})}\BibitemShut {NoStop}%
\bibitem [{\citenamefont {Kawashima}(2004)}]{Kawashima2004}%
  \BibitemOpen
  \bibfield  {author} {\bibinfo {author} {\bibfnamefont {N.}~\bibnamefont
  {Kawashima}},\ }\href {https://doi.org/10.1143/JPSJ.73.3219} {\bibfield
  {journal} {\bibinfo  {journal} {J. Phys. Soc. Jpn.}\ }\textbf {\bibinfo
  {volume} {73}},\ \bibinfo {pages} {3219} (\bibinfo {year}
  {2004})}\BibitemShut {NoStop}%
\bibitem [{\citenamefont {H\"alg}\ \emph
  {et~al.}(2015{\natexlab{a}})\citenamefont {H\"alg}, \citenamefont
  {H\"uvonen}, \citenamefont {Guidi}, \citenamefont {Quintero-Castro},
  \citenamefont {Boehm}, \citenamefont {Regnault}, \citenamefont {Hagiwara},\
  and\ \citenamefont {Zheludev}}]{HaelgHuvonenGuidi2015}%
  \BibitemOpen
  \bibfield  {author} {\bibinfo {author} {\bibfnamefont {M.}~\bibnamefont
  {H\"alg}}, \bibinfo {author} {\bibfnamefont {D.}~\bibnamefont {H\"uvonen}},
  \bibinfo {author} {\bibfnamefont {T.}~\bibnamefont {Guidi}}, \bibinfo
  {author} {\bibfnamefont {D.~L.}\ \bibnamefont {Quintero-Castro}}, \bibinfo
  {author} {\bibfnamefont {M.}~\bibnamefont {Boehm}}, \bibinfo {author}
  {\bibfnamefont {L.~P.}\ \bibnamefont {Regnault}}, \bibinfo {author}
  {\bibfnamefont {M.}~\bibnamefont {Hagiwara}},\ and\ \bibinfo {author}
  {\bibfnamefont {A.}~\bibnamefont {Zheludev}},\ }\href
  {https://doi.org/10.1103/PhysRevB.92.014412} {\bibfield  {journal} {\bibinfo
  {journal} {Phys. Rev. B}\ }\textbf {\bibinfo {volume} {92}},\ \bibinfo
  {pages} {014412} (\bibinfo {year} {2015}{\natexlab{a}})}\BibitemShut
  {NoStop}%
\bibitem [{\citenamefont {Povarov}\ \emph
  {et~al.}(2015{\natexlab{b}})\citenamefont {Povarov}, \citenamefont
  {Schmidiger}, \citenamefont {Reynolds}, \citenamefont {Bewley},\ and\
  \citenamefont {Zheludev}}]{PovarovSchmidigerReynolds2015}%
  \BibitemOpen
  \bibfield  {author} {\bibinfo {author} {\bibfnamefont {K.~Y.}\ \bibnamefont
  {Povarov}}, \bibinfo {author} {\bibfnamefont {D.}~\bibnamefont {Schmidiger}},
  \bibinfo {author} {\bibfnamefont {N.}~\bibnamefont {Reynolds}}, \bibinfo
  {author} {\bibfnamefont {R.}~\bibnamefont {Bewley}},\ and\ \bibinfo {author}
  {\bibfnamefont {A.}~\bibnamefont {Zheludev}},\ }\href
  {https://doi.org/10.1103/PhysRevB.91.020406} {\bibfield  {journal} {\bibinfo
  {journal} {Phys. Rev. B}\ }\textbf {\bibinfo {volume} {91}},\ \bibinfo
  {pages} {020406(R)} (\bibinfo {year} {2015}{\natexlab{b}})}\BibitemShut
  {NoStop}%
\bibitem [{\citenamefont {H\"alg}\ \emph
  {et~al.}(2015{\natexlab{b}})\citenamefont {H\"alg}, \citenamefont
  {H\"uvonen}, \citenamefont {Butch}, \citenamefont {Demmel},\ and\
  \citenamefont {Zheludev}}]{HaelgHuvonenButch2015}%
  \BibitemOpen
  \bibfield  {author} {\bibinfo {author} {\bibfnamefont {M.}~\bibnamefont
  {H\"alg}}, \bibinfo {author} {\bibfnamefont {D.}~\bibnamefont {H\"uvonen}},
  \bibinfo {author} {\bibfnamefont {N.~P.}\ \bibnamefont {Butch}}, \bibinfo
  {author} {\bibfnamefont {F.}~\bibnamefont {Demmel}},\ and\ \bibinfo {author}
  {\bibfnamefont {A.}~\bibnamefont {Zheludev}},\ }\href
  {https://doi.org/10.1103/PhysRevB.92.104416} {\bibfield  {journal} {\bibinfo
  {journal} {Phys. Rev. B}\ }\textbf {\bibinfo {volume} {92}},\ \bibinfo
  {pages} {104416} (\bibinfo {year} {2015}{\natexlab{b}})}\BibitemShut
  {NoStop}%
\bibitem [{\citenamefont {Lake}\ \emph {et~al.}(2005)\citenamefont {Lake},
  \citenamefont {Tennant}, \citenamefont {Frost},\ and\ \citenamefont
  {Nagler}}]{LakeTennantFrost2005}%
  \BibitemOpen
  \bibfield  {author} {\bibinfo {author} {\bibfnamefont {B.}~\bibnamefont
  {Lake}}, \bibinfo {author} {\bibfnamefont {D.~A.}\ \bibnamefont {Tennant}},
  \bibinfo {author} {\bibfnamefont {C.~D.}\ \bibnamefont {Frost}},\ and\
  \bibinfo {author} {\bibfnamefont {S.~E.}\ \bibnamefont {Nagler}},\ }\href
  {https://doi.org/10.1038/nmat1327} {\bibfield  {journal} {\bibinfo  {journal}
  {Nature Mater.}\ }\textbf {\bibinfo {volume} {4}},\ \bibinfo {pages} {329}
  (\bibinfo {year} {2005})}\BibitemShut {NoStop}%
\bibitem [{\citenamefont {Jeong}\ and\ \citenamefont
  {R\o{}nnow}(2015)}]{JeongRonnow2015}%
  \BibitemOpen
  \bibfield  {author} {\bibinfo {author} {\bibfnamefont {M.}~\bibnamefont
  {Jeong}}\ and\ \bibinfo {author} {\bibfnamefont {H.~M.}\ \bibnamefont
  {R\o{}nnow}},\ }\href {https://doi.org/10.1103/PhysRevB.92.180409} {\bibfield
   {journal} {\bibinfo  {journal} {Phys. Rev. B}\ }\textbf {\bibinfo {volume}
  {92}},\ \bibinfo {pages} {180409(R)} (\bibinfo {year} {2015})}\BibitemShut
  {NoStop}%
\bibitem [{\citenamefont {Orlova}\ \emph {et~al.}(2017)\citenamefont {Orlova},
  \citenamefont {Blinder}, \citenamefont {Kermarrec}, \citenamefont {Dupont},
  \citenamefont {Laflorencie}, \citenamefont {Capponi}, \citenamefont
  {Mayaffre}, \citenamefont {Berthier}, \citenamefont {Paduan-Filho},\ and\
  \citenamefont {Horvati\ifmmode~\acute{c}\else
  \'{c}\fi{}}}]{OrlovaBlinderKermarrec2017}%
  \BibitemOpen
  \bibfield  {author} {\bibinfo {author} {\bibfnamefont {A.}~\bibnamefont
  {Orlova}}, \bibinfo {author} {\bibfnamefont {R.}~\bibnamefont {Blinder}},
  \bibinfo {author} {\bibfnamefont {E.}~\bibnamefont {Kermarrec}}, \bibinfo
  {author} {\bibfnamefont {M.}~\bibnamefont {Dupont}}, \bibinfo {author}
  {\bibfnamefont {N.}~\bibnamefont {Laflorencie}}, \bibinfo {author}
  {\bibfnamefont {S.}~\bibnamefont {Capponi}}, \bibinfo {author} {\bibfnamefont
  {H.}~\bibnamefont {Mayaffre}}, \bibinfo {author} {\bibfnamefont
  {C.}~\bibnamefont {Berthier}}, \bibinfo {author} {\bibfnamefont
  {A.}~\bibnamefont {Paduan-Filho}},\ and\ \bibinfo {author} {\bibfnamefont
  {M.}~\bibnamefont {Horvati\ifmmode~\acute{c}\else \'{c}\fi{}}},\ }\href
  {https://doi.org/10.1103/PhysRevLett.118.067203} {\bibfield  {journal}
  {\bibinfo  {journal} {Phys. Rev. Lett.}\ }\textbf {\bibinfo {volume} {118}},\
  \bibinfo {pages} {067203} (\bibinfo {year} {2017})}\BibitemShut {NoStop}%
\bibitem [{\citenamefont {Dupont}\ \emph {et~al.}(2017)\citenamefont {Dupont},
  \citenamefont {Capponi},\ and\ \citenamefont
  {Laflorencie}}]{DupontCapponiLaflorencie2017}%
  \BibitemOpen
  \bibfield  {author} {\bibinfo {author} {\bibfnamefont {M.}~\bibnamefont
  {Dupont}}, \bibinfo {author} {\bibfnamefont {S.}~\bibnamefont {Capponi}},\
  and\ \bibinfo {author} {\bibfnamefont {N.}~\bibnamefont {Laflorencie}},\
  }\href {https://doi.org/10.1103/PhysRevLett.118.067204} {\bibfield  {journal}
  {\bibinfo  {journal} {Phys. Rev. Lett.}\ }\textbf {\bibinfo {volume} {118}},\
  \bibinfo {pages} {067204} (\bibinfo {year} {2017})}\BibitemShut {NoStop}%
\end{thebibliography}%

\end{document}